\documentclass[draft,noendmarks]{csreport}

\pdfoutput=1

\usepackage[UKenglish]{babel}

\newif\iffull
\fulltrue

\usetikzlibrary{calc}
\tikzset{play/.style={circle,draw,minimum size=#1}}
\tikzset{play/.default=0.625cm}
\tikzset{prob/.style={diamond,draw,minimum size=#1}}
\tikzset{prob/.default=0.7cm}
\tikzset{end/.style={rectangle,draw,minimum size=#1}}
\tikzset{end/.default=0.55cm}
\tikzset{every label/.style={font=\small}}
\tikzset{action/.style={font=\footnotesize}}
\tikzset{every node/.style={font=\small}}

\newcommand{\NE}{\mbox{NE}\xspace}
\newcommand{\PosNE}{\mbox{PosNE}\xspace}
\newcommand{\StatNE}{\mbox{StatNE}\xspace}

\newcommand{\PureNE}{\mbox{PureNE}\xspace}

\newcommand{\SAT}{\mbox{SAT}\xspace}
\newcommand{\SqrtSum}{\mbox{SqrtSum}\xspace}
\newcommand{\true}{\mbox{true}}
\newcommand{\false}{\mbox{false}}

\newcommand{\eg}{e.g.\@\xspace}
\newcommand{\ie}{i.e.\@\xspace}
\renewcommand{\iff}{if and only if\xspace}
\newcommand{\wrt}{w.r.t.\@\xspace}
\newcommand{\wlg}{w.l.o.g.\@\xspace}
\newcommand{\pl}[1]{player~$#1$\xspace}
\newcommand{\Pl}[1]{Player~$#1$\xspace}
\newcommand{\pli}{\pl{i}}

\newenvironment{map}{\begin{array}{@{}r@{\colon}l@{}}}{\end{array}}

\DeclareMathOperator{\init}{init}
\DeclareMathOperator{\inc}{inc}
\DeclareMathOperator{\dec}{dec}
\DeclareMathOperator{\zero}{zero}

\DeclareMathOperator{\Prob}{Pr}

\DeclareMathOperator{\Exp}{E}
\newcommand{\diff}{\mathrm{d}}

\let\pay\origphi
\DeclareMathOperator{\pval}{pval}
\newcommand\One{\mathbb{1}}
\let\upsigma\sigma
\newcommand{\Ind}{{[k]}}
\DeclareMathOperator{\In}{In}
\DeclareMathOperator{\Out}{Out}

\makeatletter
\newcommand{\newreptheorem}[2]{%
 \newtheorem*{rep@#1}{\rep@title}
 \newenvironment{rep#1}[1]{%
  \def\rep@title{#2 \ref*{##1}}%
  \begin{rep@#1}}%
  {\end{rep@#1}%
 }%
}
\makeatother

\theoremstyle{definition}
\newreptheorem{theorem}{Theorem}

\title{The Complexity of Nash Equilibria\\ in Limit-Average
Games\thanks{This work was supported by
ESF RNP ``Games for Design and Verification'' (GAMES),
the French project ANR-06-SETI-003 (DOTS) and
EPSRC grant EP/G050112/1.}}
\author{Michael Ummels\addr{1} and Dominik Wojtczak\addr{2,3}}
\address{LSV, CNRS \& ENS Cachan, France \\
\email{ummels@lsv.ens-cachan.fr}}
\address{University of Liverpool, UK \\
\email{d.k.wojtczak@liv.ac.uk}}
\address{Oxford University Computing Laboratory, UK}

\begin{document}

\maketitle

\begin{abstract}
We study the computational complexity of Nash equilibria in
concurrent games with limit-average objectives.
In particular, we prove that the existence of a Nash equilibrium
in randomised strategies is undecidable, while the existence of
a Nash equilibrium in pure strategies is decidable, even if we put
a~constraint on the payoff of the equilibrium. Our undecidability
result holds even for a restricted class of concurrent games, where
nonzero rewards occur only on terminal states. Moreover, we show
that the constrained existence problem is undecidable not only for
concurrent games but for turn-based games with the same restriction
on rewards. Finally, we prove that the constrained existence
problem for Nash equilibria in (pure or randomised) stationary
strategies is decidable and analyse its complexity.
\end{abstract}

\section{Introduction}

Concurrent games provide a versatile model for the
interaction of several components in a distributed system where the
components perform actions in parallel~\cite{Henzinger05}.
Classically, such a system is modelled by a family of concurrent two-player
games, one for each component, where one component tries to fulfil its
specification against the coalition of all other components. In practice,
this modelling is often too pessimistic because it ignores the specifications
of the other components. We argue that a distributed system is more faithfully
modelled by a multiplayer game where each player has her own objective, which
is independent of the other players' objectives.

Another objection to the classical theory of verification and synthesis
has been that specifications are \emph{qualitative}: either the
specification is fulfilled, or it is violated. Examples of such
specifications include reachability properties, where a certain set of
target states has to be reached, or safety properties, where a certain
set of states has to be avoided.
In practice, many specifications are of a \emph{quantitative}
nature, examples of which include minimising average power consumption or
maximising average throughput. Specifications of the latter kind can be
expressed by assigning (positive or negative) rewards to states or
transitions and considering
the \emph{limit-average} reward gained from an infinite play. In fact,
concurrent games where a player's payoff is defined in such a way have been
a central topic in game theory (see the related work section below).

The most common solution concept for games with multiple players is that of a
Nash equilibrium \cite{Nash50}. In a Nash equilibrium, no player can improve
her payoff by changing her strategy unilaterally.
Unfortunately, Nash equilibria do not always exist in concurrent games, and
if they exist, they may not be unique.
In~applications, one might look for an equilibrium where some players
receive a high payoff while other players receive a low payoff.
Formulated as a decision problem, given a game with $k$~players
and thresholds $\vec{x},\vec{y}\in(\bbQ\cup\{\pm\infty\})^k$, we~want to
know whether the game has a Nash equilibrium whose payoff
lies in-between $\vec{x}$ and~$\vec{y}$; we call this decision problem \NE.

The problem \NE comes in several variants, depending on the type of strategies
one considers: On the one hand, strategies may be \emph{randomised} (allowing
randomisation over actions) or \emph{pure} (not allowing such randomisation).
\iffull
On~the other hand, one can restrict to strategies that use finite memory or
even to \emph{stationary} strategies,
\fi
which only depend on the last state.
Indeed, we show that these restrictions give rise to distinct decision
problems, which have to be analysed separately.

Our results show that the complexity of \NE highly depends
on the type of strategies that realise the equilibrium. In particular, we prove
the following results, which yield an almost complete picture of the complexity
of~\NE:
\begin{enumerate}
\item \NE for pure stationary strategies
\iffull
(or pure strategies with bounded memory)
\fi
is \NP-complete.
\item \NE for stationary strategies
\iffull
(or randomised strategies with bounded memory)
\fi
is decidable in \PSpace,
but hard for both \NP and \SqrtSum.
\item \NE for arbitrary pure strategies is \NP-complete.
\item \NE for arbitrary randomised strategies is undecidable.
\end{enumerate}

All of our lower bounds for \NE and, in particular, our undecidability result
hold already for a subclass of concurrent games where Nash equilibria are
guaranteed to exist, namely for \emph{turn-based} games.
If this assumption is relaxed and Nash
equilibria are not guaranteed to exist, we prove that
even the plain existence problem for Nash equilibria is undecidable.
Moreover, many of our lower bounds hold already for games where
non-zero rewards only occur on terminal states, and thus also for games
where each player wants to maximise the \emph{total sum} of the rewards.

As a byproduct of our decidability proof for pure strategies,
we give a poly\-nomial-time algorithm for deciding whether in a multi-weighted
graph there exists a path whose limit-average weight vector lies between
two given thresholds, a result that is of independent interest. For
instance, our algorithm can be used for deciding the emptiness of a
\emph{multi-threshold mean-payoff language}~\cite{AlurDMW09} in
polynomial time.

\subsubsection{Related work}

Concurrent and, more generally, stochastic games go back to
\citet{Shapley53}, who proved the existence of the \emph{value} for
\emph{discounted two-player zero-sum} games. This result was later
generalised by \citet{Fink64} who proved that every discounted
game has a Nash equilibrium. \Citet{Gillette57} introduced limit-average
objectives, and \citet{MertensN81} proved the existence of
the value for stochastic two-player zero-sum games with
limit-average objectives.
Unfortunately, as demonstrated by \citet{Everett57},
these games do, in~general, not admit a Nash equilibrium (see
\cref{ex:hide-or-run}).
However, \citet{Vielle00a,Vielle00b} proved that,
for all $\epsilon>0$, every two-player stochastic limit-average game admits
an \emph{$\epsilon$-equilibrium}, \ie a pair of strategies where each
player can gain at most~$\epsilon$ from switching her strategy. Whether
such equilibria always exist in games with more than two players
is an important open question~\cite{NeymanS03}.

Determining the complexity of Nash equilibria has attracted much
interest in recent years. In particular, a series of papers culminated in the
result that computing a Nash equilibrium of a finite two-player game in
\emph{strategic form} is complete for the complexity class \PPAD
\cite{ChenDT09,DaskalakisGP09}. The constrained existence problem,
where one looks for a Nash equilibrium with certain properties, has also
been investigated for games in strategic form. In particular,
\citet{ConitzerS03} showed that deciding whether there exists a Nash
equilibrium whose payoff exceeds a given threshold and
related decision problems are \NP-complete
for two-player games in strategic form.

For concurrent games with limit-average objectives, most algorithmic results
concern two-player zero-sum games.
In the turn-based case, these games are commonly known as
\emph{mean-payoff games} \cite{EhrenfeuchtM79,ZwickP96}. While it is known that
the value of such a game can be computed in pseudo-polynomial time, it is still
open whether there exists a polynomial-time algorithm for solving mean-payoff
games. A related model are \emph{multi-dimensional mean-payoff games}
where one player tries to maximise several mean-payoff conditions at the
same time~\cite{ChatterjeeDHR10}. In particular, \citet{VelnerR11} showed
that the value problem for these games is \coNP-complete.

One subclass of concurrent games with limit-average objectives that
has been studied in the multiplayer setting are concurrent games
with reachability objectives. In particular, \citet{BouyerBM10b}
showed that the constrained existence problem for Nash equilibria
is \NP-complete for these games (see also \cite{Ummels08,FismanKL10}).
We extend their result to limit-average objectives. However, we assume
that strategies can observe actions (a common assumption in game theory),
which they do not. Hence, while
our result is more general \wrt the type of objectives we consider, their
result is more general \wrt the type of strategies they allow.

In a recent paper~\cite{UmmelsW09}, we studied the complexity of Nash
equilibria in \emph{stochastic} games with reachability objectives.
In particular, we proved that \NE for pure strategies is undecidable
in this setting. Since we prove here that this problem is decidable
in the non-stochastic setting, this undecidability result can be
explained by the presence of probabilistic transitions in stochastic
games. On~the other hand, we prove in this paper that randomisation in
strategies also leads to undecidability, a question that was left open
in~\cite{UmmelsW09}.

\section{Concurrent Games}
\label{sect:games}

Concurrent games are played by finitely many players on a finite
state space. Formally, a concurrent game is given by
\begin{itemize}
 \item a finite nonempty set~$\Pi$ of \emph{players},
\eg $\Pi=\{0,1,\dots,k-1\}$,
 \item a finite nonempty set~$S$ of \emph{states},
 \item for each \pli and each state~$s$ a nonempty set $\Gamma_i(s)$ of
\emph{actions} taken from a finite set~$\Gamma$,
 \item a \emph{transition function} $\delta\colon S\times\Gamma^\Pi\to S$,
 \item for each player $i\in\Pi$ a \emph{reward function} $r_i\colon S\to\bbR$.
\end{itemize}
For computational purposes, we assume that all rewards are rational numbers
with numerator and denominator given in binary.
We say that an action profile $\vec{a}=(a_i)_{i\in\Pi}$ is \emph{legal}
at state~$s$ if $a_i\in\Gamma_i(s)$ for each $i\in\Pi$.
Finally, we call a state~$s$ \emph{controlled} by \pli if
$\abs{\Gamma_j(s)}=1$ for all $j\neq i$, and we say that a game is
\emph{turn-based} if each state is controlled by (at least) one player.
For turn-based games, an action of the controlling player prescribes to go to
a certain state. Hence, we will usually omit actions in turn-based games.


For a tuple $\vec{x}=(x_i)_{i\in\Pi}$, where the elements~$x_i$ belong to an
arbitrary set~$X$, and an element $x\in X$, we denote by $\vec{x}_{-i}$ the
restriction of~$\vec{x}$ to $\Pi\setminus\{i\}$ and by $(\vec{x}_{-i},x)$
the unique tuple $\vec{y}\in X^\Pi$ with $y_i=x$ and
$\vec{y}_{-i}=\vec{x}_{-i}$.

A play of a game~$\calG$ is an infinite sequence
$s_0\vec{a}_0s_1\vec{a}_1\ldots\in{(S\cdot\Gamma^\Pi)}^\omega$
such that $\delta(s_j,\vec{a}_j)=s_{j+1}$ for all $j\in\bbN$.
For each player, a play $\pi=s_0\vec{a}_0s_1\vec{a}_1\ldots$
gives rise to an infinite sequence of rewards. There are different criteria to
evaluate this sequence and map it to a \emph{payoff}. In this paper,
we consider the \emph{limit-average} (or~\emph{mean-payoff})
criterion, where the payoff of~$\pi$ for \pli is defined by
\iffull
\[ \pay_i(\pi)\coloneq \liminf_{n\to\infty}
\,\frac{1}{n}\sum_{j=0}^{n-1} r_i(s_j).\]
\fi
Note that this payoff mapping is \emph{prefix-independent}, \ie
$\pay_i(\pi)=\pay_i(\pi')$ if $\pi'$~is a suffix of~$\pi$.
An important special case are games where non-zero rewards occur only on
\emph {terminal} states, \ie states~$s$ with
$\delta(s,\vec{a})=s$
for all (legal) $\vec{a}\in\Gamma^\Pi$. These games were introduced by
\citet{Everett57} under the name \emph{recursive games}, but we prefer to
call them \emph{terminal-reward games}. Hence, in a terminal-reward
game, $\pay_i(\pi)=r_i(s)$ if $\pi$~enters a terminal state~$s$
and $\pay_i(\pi)=0$ otherwise.

Often, it is convenient to designate an \emph{initial} state.
An \emph{initialised} game is thus a tuple $(\calG,s_0)$ where $\calG$ is a
concurrent game and $s_0$~is one of its states.

\subsubsection{Strategies and strategy profiles}

For a finite set~$X$, we denote by $\calD(X)$ the set of probability
distributions over~$X$.
A \emph{(randomised) strategy} for \pli in~$\calG$ is a mapping
$\sigma\colon{(S\cdot\Gamma^\Pi)}^*\cdot S\to\calD(\Gamma)$ assigning to each
possible \emph{history} ${xs\in{(S\cdot\Gamma^\Pi)}^*\cdot S}$
a probability distribution~$\sigma(xs)$ over actions such that
$\sigma(xs)(a)>0$ only if $a\in\Gamma_i(s)$.
We write $\sigma(a\mid xs)$ for the probability assigned to
$a\in\Gamma$ by the distribution $\sigma(xs)$.
A \emph{(randomised) strategy profile of $\calG$} is a tuple
$\vec{\sigma}=(\sigma_i)_{i\in\Pi}$ of strategies in~$\calG$, one
for each player.
\iffull
Note that a strategy profile can be identified
with a function $\vec{\sigma}\colon{(S\cdot\Gamma^\Pi)}^*\cdot S\to
\calD(\Gamma)^\Pi$.
\fi

A strategy~$\sigma$ for \pli is called \emph{pure} if for each
$xs\in{(S\cdot\Gamma^\Pi)}^*\cdot S$ the distribution $\sigma(xs)$
is \emph{degenerate}, \ie there exists $a\in\Gamma_i(s)$ with
$\sigma(a\mid xs)=1$.
Note that a pure strategy can be identified with a function
$\sigma\colon{(S\cdot\Gamma^\Pi)}^*\cdot S\to\Gamma$.
A strategy profile $\vec{\sigma}=(\sigma_i)_{i\in\Pi}$
is called \emph{pure} if each $\sigma_i$~is
pure, in which case we can identify $\vec{\sigma}$ with a mapping
${(S\cdot\Gamma^\Pi)}^*\cdot S\to\Gamma^\Pi$.
Note that, given an initial state~$s_0$ and a pure strategy
profile~$\vec{\sigma}$,
there exists a unique play $\pi=s_0\vec{a}_0s_1\vec{a}_1\ldots$
such that $\vec{\sigma}(s_0\vec{a}_0\dots\vec{a}_{j-1}s_j)
=\vec{a}_j$ for all $j\in\bbN$; we call~$\pi$ the play \emph{induced}
by~$\vec{\sigma}$ from~$s_0$.

\iffull
A \emph{memory structure} for~$\calG$ is a triple $\frakM=(M,\delta,m_0)$,
where $M$~is a set of \emph{memory states},
$\delta\colon M\times S\times\Gamma^\Pi\to M$ is the
\emph{update function}, and $m_0\in M$ is the \emph{initial memory}.
A (\emph{randomised}) \emph{strategy with memory~$\frakM$} for \pli is a
function $\sigma\colon M\times S\to\calD(\Gamma)$
such that $\sigma(m,s)(a)>0$ only if $a\in\Gamma_i(s)$.
The strategy~$\sigma$ is \emph{pure} if the distribution~$\sigma(m,s)$
is degenerate for all $m\in M$ and $s\in S$. A~(pure)
strategy~$\sigma$ with memory~$\frakM$ can be viewed as a (pure)
strategy~$\sigma'$ in the usual sense by setting
$\sigma'(xs)=\sigma(\delta^*(x),s)$, where
$\delta^*(x)$ is defined inductively by $\delta^*(\epsilon)=m_0$ and
$\delta^*(x\cdot s\vec{a})=\delta(\delta^*(x),s,\vec{a})$.
A \emph{finite-state} strategy is a strategy~$\sigma$ with finite
memory~$\frakM$. If the memory~$\frakM$ is a singleton, we
call~$\sigma$ \emph{stationary}. Moreover, we call a strategy
\emph{positional} if it is both pure and stationary.
A stationary strategy can thus be represented by a mapping
$\sigma\colon S\to\calD(\Gamma)$, and a positional strategy by a
mapping $\sigma\colon S\to\Gamma$.
Finally, we call a strategy
profile finite-state, stationary or positional if each strategy
in the profile has the respective property.
\fi

\subsubsection{The probability measure induced by a strategy profile}

Given an initial state $s_0\in S$ and a strategy profile
$\vec{\sigma}=(\sigma_i)_{i\in\Pi}$, the \emph{conditional probability}
of $\vec{a}\in\Gamma^\Pi$ given the history
$xs\in{(S\cdot\Gamma^\Pi)}^*\cdot S$ equals
\iffull
\[
\vec{\sigma}(\vec{a}\mid xs)\coloneq\prod_{i\in\Pi}\sigma_i(a_i\mid xs).
\]
\fi
The probabilities $\vec{\sigma}(\vec{a}\mid xs)$ induce a probability measure
on the Borel $\sigma$-algebra over ${(S\cdot\Gamma^\Pi)}^\omega$ as
follows: The probability of a basic open set
$s_1\vec{a}_1\dots s_n\vec{a}_n\cdot{(S\cdot\Gamma^\Pi)}^\omega$
equals the product
$\prod_{j=1}^n\,\vec{\sigma}(\vec{a}_j\mid s_1\vec{a}_1\ldots\vec{a}_{j-1}s_j)$
if $s_1=s_0$ and $\delta(s_j,\vec{a}_j)=s_{j+1}$ for all $1\leq j<n$;
in all other cases, this probability is~$0$. By \emph{Carath\'eodory's
extension theorem}, this extends to a
unique probability measure assigning a probability to every Borel subset of
${(S\cdot\Gamma^\Pi)}^\omega$, which we denote by~$\Prob_{s_0}^{\vec{\sigma}}$.
Via the natural projection ${(S\cdot\Gamma^\Pi)}^\omega\to S^\omega$, we
obtain a probability measure on the Borel $\upsigma$-algebra over $S^\omega$.
We abuse notation and denote this measure also by~$\Prob_{s_0}^{\vec{\sigma}}$;
it~should always be clear from the context to which measure we are referring
to.
Finally, we denote by $\Exp_{s_0}^{\vec{\sigma}}$ the expectation operator
that corresponds to $\Prob_{s_0}^{\vec{\sigma}}$, \ie
$\Exp_{s_0}^{\vec{\sigma}}(f)=\int\! f\,\diff\!\Prob_{s_0}^{\vec{\sigma}}$
for all Borel measurable functions
$f\colon{(S\cdot\Gamma^\Pi)}^\omega\to\bbR\cup\{\pm\infty\}$ or
$f\colon S^\omega\to\bbR\cup\{\pm\infty\}$.
In particular, we are interested
in the quantities $p_i\coloneq\Exp_{s_0}^{\vec{\sigma}}(\pay_i)$. We
call~$p_i$ the \emph{(expected) payoff} of~$\vec{\sigma}$ for \pli and the
vector $(p_i)_{i\in\Pi}$ the (expected) payoff of~$\vec{\sigma}$.
\iffull
Finally,
we call a history $x\in{(S\cdot\Gamma^\Pi)}^*\cdot S$ \emph{consistent}
with~$\vec{\sigma}$ if $\Prob^{\vec{\sigma}}_{s_0}
(x\cdot{(S\cdot\Gamma^\Pi)}^\omega)>0$.
\fi

\iffull
In order to apply known results about Markov chains, we can also view the
stochastic process induced by a strategy profile~$\vec{\sigma}$ as a
countable Markov chain~$\calG^{\vec{\sigma}}$, defined as follows:
The set of states of $\calG^{\vec{\sigma}}$ equals the set
${(S\cdot\Gamma^\Pi)}^*\cdot S$ of histories of~$\calG$. The only
transitions from a state~$xs$ lead to states of the form $xs\vec{a}t$
where $t=\delta(s,\vec{a})$, and such a transition occurs with
probability $\vec{\sigma}(\vec{a}\mid xs)$.

For each \pli, the Markov decision process~$\calG^{\vec{\sigma}_{-i}}$ has
the same states as~$\calG^{\vec{\sigma}}$, and there is a transition
from a state~$xs$ to a state $xs\vec{a}t$ with action $a\in\Gamma_i(s)$
and probability~$p$ if $a_i=a$, $\delta(s,\vec{a})=t$ and
$p=\prod_{j\neq i}\sigma_j(a_j)$. Finally, the reward of a state~$xs$
in $\calG^{\vec{\sigma}_{-i}}$ equals the reward $r_i(s)$ of the state~$s$
for \pli in~$\calG$.

If $\vec{\sigma}$~is a strategy profile with finite memory~$\frakM$,
we make $\calG^{\vec{\sigma}}$ and~$\calG^{\vec{\sigma}_{-i}}$ finite
by quotienting the state space
\wrt the equivalence relation~${\sim}$, defined by $xs\sim yt$ if $s=t$ and
$\delta^*(x)=\delta^*(y)$. In particular, if $\vec{\sigma}$ is
stationary, then the state spaces of $\calG^{\vec{\sigma}}$
and~$\calG^{\vec{\sigma}_{-i}}$ coincide with the state space of~$\calG$.
\fi

\subsubsection{Drawing concurrent games}

When drawing a concurrent game as a graph, we will adhere to the following
conventions:
States are usually depicted as circles, but terminal states are depicted as
squares. The initial state is marked by a dangling incoming edge. An edge from
$s$ to~$t$ with label~$\vec{a}$ means that $\delta(s,\vec{a})=t$ and
that $\vec{a}$~is legal at~$s$. However, the label~$\vec{a}$
might be omitted if it is not essential. In~turn-based games, the
player who controls a state is indicated by the label next to it.
Finally, a label of the form $i\colon x$ next to state~$s$ indicates
that $r_i(s)=x$; if~this reward is~$0$, the label will usually be omitted.

\section{Nash Equilibria}
\label{sect:nash}

To capture rational behaviour of selfish players, Nash \citep{Nash50}
introduced the notion of\,---\,what is now called\,---\,a \emph{Nash
equilibrium}.
Formally, given a game~$\calG$ and an initial state~$s_0$,
a strategy~$\tau$ for \pli is a \emph{best response} to a
strategy profile~$\vec{\sigma}$ if $\tau$~maximises the
expected payoff for \pli, \ie
\[\Exp_{s_0}^{\vec{\sigma}_{-i},\tau'}(\pay_i)
\leq\Exp_{s_0}^{\vec{\sigma}_{-i},\tau}(\pay_i)\]
for all strategies~$\tau'$ for \pli.
A strategy profile $\vec{\sigma}=(\sigma_i)_{i\in\Pi}$ is a \emph{Nash
equilibrium} of $(\calG,s_0)$ if for each \pli the strategy~$\sigma_i$
is a best response to~$\vec{\sigma}$. Hence, in a Nash equilibrium no player
can improve her payoff by (unilaterally) switching to a different strategy.
As the following examples demonstrate, Nash equilibria are not guaranteed to
exist in concurrent games.

\begin{example}\label{ex:hide-or-run}
Consider the terminal-reward game~$\calG_1$ depicted in \cref{fig:no-ne}
and played by players $1$ and~$2$,
which was originally presented in~\cite{AlfaroHK07}.
\begin{figure}
\centering
\begin{floatrow}
\ffigbox[.45\textwidth]{%
\begin{tikzpicture}[x=2cm,y=1cm,->]
\node (limbo) at (0,0) [play] {$s_1$};
\node (safe) at (1,0) [end,label={above:$1\colon 1$},label={below:$2\colon -1$}] {};
\node (dead) at (-1,0) [end,label={above:$1\colon -1$},label={below:$2\colon 1$}] {};

\draw (0,0.8) to (limbo);
\draw (limbo) to node[above,action] {$(a,b)$} node[below,action] {$(b,a)$} (safe);
\draw (limbo) to node[below,action] {$(b,b)$} (dead);
\draw[loop,out=-45,in=-135,looseness=6] (limbo) to node[below,action] {$(a,a)$} (limbo);
\end{tikzpicture}}
{\caption{\label{fig:no-ne}A terminal-reward game that has no Nash equilibrium}}
\ffigbox[.45\textwidth]{%
\begin{tikzpicture}[x=2cm,y=1cm,->]
\node (limbo) at (0,0) [play,label={above left:$2\colon 1\!\!$}] {$s_1$};
\node (safe) at (1,0) [end,label={above:$1\colon 1$}] {};
\node (dead) at (-1,0) [end,label={above:$2\colon 1$}] {};
\draw (limbo) to node[above,action] {$(a,b)$} node[below,action] {$(b,a)$} (safe);
\draw (0,0.8) to (limbo);

\draw (limbo) to node[below,action] {$(b,b)$} (dead);
\draw[loop,out=-45,in=-135,looseness=6] (limbo) to node[below,action] {$(a,a)$} (limbo);
\end{tikzpicture}}
{\caption{\label{fig:no-ne-2}A limit-average game that has no Nash equilibrium}}
\end{floatrow}
\end{figure}
We claim that $(\calG_1,s_1)$ does not have a Nash
equilibrium. First note that, for each $\epsilon>0$, \pl1 can ensure a
payoff of $1-2\epsilon$ by the stationary strategy that selects action~$b$
with probability~$\epsilon$. Hence, every Nash equilibrium
$(\sigma,\tau)$ of $(\calG_1,s_1)$ must have payoff $(1,-1)$.
Now we distinguish whether $\sigma(b\mid (s_1(a,a))^k s_1)=0$ for
all $k\in\bbN$ or not. In the first case, there must exist $k\in\bbN$
such that
$\tau(b\mid (s_1(a,a))^k s_1)>0$ (otherwise $(\sigma,\tau)$ would
not have payoff $(1,-1)$). But then \Pl2 can improve her payoff
by always playing action~$a$ with probability~$1$, a~contradiction
to $(\sigma,\tau)$ being a Nash equilibrium.
In the second case, consider the least~$k$ such that
$p\coloneq\sigma(b\mid (s_1(a,a))^k s_1)>0$.
By choosing action~$b$ with probability~$1$ for the history
$(s_1(a,a))^k s_1$ and choosing action~$a$ with probability~$1$ for
all other histories, \pl2 can ensure payoff~$p$, again a~contradiction
to $(\sigma,\tau)$ being a Nash equilibrium.
\end{example}

\begin{example}\label{ex:hide-or-run-2}
A variation of the previous game is the game~$\calG_2$, which is depicted
in \cref{fig:no-ne-2} and also played by players $1$ and~$2$. It is not
a terminal-reward game, but the only rewards that occur in the game are
$0$ and~$1$. Using almost the same argumentation as in \cref{ex:hide-or-run},
we can show that $(\calG_2,s_1)$ has no Nash equilibrium either.
\end{example}

It follows from Nash's theorem \citep{Nash50} that every game
whose arena is a tree (or~a DAG) has a Nash equilibrium.
Another important special case of concurrent limit-average games where Nash
equilibria always exist are turn-based games. For these games,
\citet{ThuijsmanR97} proved not only the
existence of arbitrary Nash equilibria but of pure finite-state ones.

To measure the complexity of Nash equilibria in concurrent games,
we~introduce the following decision problem, which we call \NE:
\begin{quote}
Given a game~$\calG$, a state~$s_0$ and thresholds
$\vec{x},\vec{y}\in(\bbQ\cup\{\pm\infty\})^\Pi$, decide
whether $(\calG,s_0)$ has a Nash equilibrium with
payoff $\geq\vec{x}$ and $\leq\vec{y}$.
\end{quote}
Note that we have not put any restriction on the type of strategies that
realise the equilibrium.
It is natural to restrict the search space to profiles of pure, stationary
or positional strategies.
These restrictions give rise to different decision
problems, which we call \PureNE, \StatNE and \PosNE, respectively.

Before we analyse the complexity of these problems, let us convince ourselves
that these problems are not just different faces of the same coin.
We first show that the decision problems
where we look for equilibria in randomised strategies are distinct
from the ones where we look for equilibria in pure strategies.

\begin{proposition}\label{prop:pure-not-enough}
There exists a turn-based terminal-reward game that has a stationary
Nash equilibrium
where \pl0 receives payoff~$1$ but that has no pure Nash equilibrium where
\pl0 receives payoff $>0$.
\end{proposition}

\begin{proof}\label{prop:no-pure-nash}
Consider the game depicted in \cref{fig:no-pure-nash} and played by three
players $0$, $1$ and~$2$.
\begin{figure}
\centering
\begin{floatrow}
\ffigbox[.525\textwidth]{%
\begin{tikzpicture}[x=1.4cm,y=1.3cm,->]
\useasboundingbox (-0.65,0.6) rectangle (3.75,-1.6);
\node (init) at (-0.7,0) {};
\node (0) at (0,0) [play,label={above:1}] {$s_0$};
\node (1) at (1,0) [play,label={above:2}] {$s_1$};
\node (2) at (2,0) [play,label={above:0}] {$s_2$};
\node (3) at (2,-1) [end,label={right:$\begin{map}0 & 1 \\ 1 & 2\end{map}$}] {};
\node (4) at (3,0) [end,label={right:$\begin{map}0 & 1 \\ 2 & 2\end{map}$}] {};
\node (5) at (0,-1) [end,label={below:$1\colon 1$}] {};
\node (6) at (1,-1) [end,label={below:$2\colon 1$}] {};

\draw (init) -- (0);
\draw (0) -- (1);
\draw (0) -- (5);
\draw (1) -- (2);
\draw (1) -- (6);
\draw (2) -- (3);
\draw (2) -- (4);
\end{tikzpicture}}%
{\caption{\label{fig:no-pure-nash}A game with no pure Nash equilibrium
where \pl0 wins with positive probability}}
\ffigbox[.415\textwidth]{%
\begin{tikzpicture}[x=1.4cm,y=1.3cm,->]
\useasboundingbox (-0.75,0.6) rectangle (2.75,-1.65);
\node (init) at (-0.7,0) {};
\node (0) at (0,0) [play,label={above:1}] {$s_0$};
\node (1) at (1,0) [play,label={above:2}] {$s_1$};
\node (2) at (2,0) [end,label={right:$0\colon 1$}] {};
\node (3) at (0,-1) [end,label={left:$1\colon 1$}] {};
\node (4) at (1,-1) [play,label={below:$0$}] {$s_2$};
\node (5) at (2,-1) [end,label={right:$2\colon 1$}] {};

\draw (init) -- (0);
\draw (0) -- (1);
\draw (0) -- (4);
\draw (1) -- (2);
\draw (1) -- (4);
\draw (4) -- (3);
\draw (4) -- (5);
\end{tikzpicture}}%
{\caption{\label{fig:no-stat-nash}A game with no stationary Nash
equilibrium where \pl0 wins with positive probability}}
\end{floatrow}
\end{figure}
Clearly, the stationary strategy profile where at state~$s_2$ \pl0
selects both outgoing transitions with probability~$\frac{1}{2}$ each,
\pl1 plays from $s_0$ to~$s_1$ and \pl2 plays from $s_1$ to~$s_2$ is
a Nash equilibrium where \pl0 receives payoff~$1$. However, in any pure
strategy profile where \pl0 receives payoff~${>0}$, either \pl1 or
\pl2 receives payoff~$0$ and could improve her payoff by switching her
strategy at $s_0$ or~$s_1$, respectively.\qed
\end{proof}

Now we show that it makes a difference whether we look for an
equilibrium in stationary strategies or not.

\begin{proposition}\label{prop:no-stat-nash}
There exists a turn-based terminal-reward game that has a pure Nash equilibrium
where \pl0 receives payoff~$1$ but that has no stationary Nash equilibrium
where \pl0 receives payoff~$>0$.
\end{proposition}

\begin{proof}
Consider the game~$\calG$ depicted in \cref{fig:no-stat-nash} and played by
three players $0$, $1$ and~$2$.
Clearly, the pure strategy profile that leads to the terminal state
with payoff~$1$ for \pl0 and where \pl0 plays ``right'' if \pl1 has
deviated and ``left'' if \pl2 has deviated is a Nash equilibrium
of $(\calG,s_0)$ with payoff~$1$ for \pl0. Now consider any stationary
equilibrium of $(\calG,s_0)$ where \pl0 receives payoff~${>0}$.
If the stationary strategy of \pl0 prescribes to play ``right''
with positive probability, then \pl2 can improve her payoff by
playing to~$s_2$ with probability~$1$, and otherwise \pl1 can
improve her payoff by playing to~$s_2$ with probability~$1$,
a contradiction.\qed
\end{proof}

It follows from \cref{prop:no-pure-nash} that \NE and \StatNE are
different from \PureNE and \PosNE, and it follows from \cref{prop:no-stat-nash}
that \NE and \PureNE are different from \StatNE and \PosNE. Hence,
all of these decision problems are pairwise distinct, and their decidability
and complexity has to be studied separately.

\section{Positional Strategies}
\label{sect:positional}

\iffull
In this section, we show that the problem \PosNE is \NP-complete;
we start by proving the upper bound.
\fi

\begin{theorem}\label{thm:posne-np}
\PosNE is in \NP.
\end{theorem}

\iffull
\begin{proof}
To decide \PosNE on input $\calG,s_0,\vec{x},\vec{y}$, we start by guessing a
positional strategy profile~$\vec{\sigma}$ of~$\calG$, \ie mappings
$\sigma_i\colon S\to\Gamma$ such that $\sigma_i(s)\in\Gamma_i(s)$ for
all $i\in\Pi$ and $s\in S$.
Then, we verify whether $\vec{\sigma}$~is a Nash equilibrium with the desired
payoff. To~do this, we first
compute the payoff~$z_i$ of~$\vec{\sigma}$ for each \pli by
computing the number $\Exp_{s_0}^{\vec{\sigma}}(\pay_i)$ in the
finite Markov chain~$\calG^{\vec{\sigma}}$.
Since $\calG^{\vec{\sigma}}$~is deterministic, this number equals the
average weight (for \pli) on the unique
simple cycle reachable from~$s_0$ and can thus be computed in
polynomial time.
Once each~$z_i$ is computed, we can
easily check whether $x_i\leq z_i\leq y_i$.
To~verify that $\vec{\sigma}$ is a Nash equilibrium, we additionally
compute, for each \pli, the value~$v_i$ of the finite
MDP~$\calG^{\vec{\sigma}_{-i}}$ from~$s_0$.
This number can be computed by identifying
the highest average weight (for \pli) on a simple cycle reachable
in~$\calG^{\vec{\sigma}_{-i}}$ from~$s_0$,
which can also be done in polynomial time \cite{Karp78}.
Clearly,
$\vec{\sigma}$ is a Nash equilibrium \iff $v_i\leq z_i$ for each \pli.\qed
\end{proof}
\fi

A result by \citet[Lemma~15]{ChatterjeeDHR10} implies that \PosNE is
\NP-hard, even for turn-based games with rewards taken from $\{-1,0,1\}$
(but with an unbounded number of players).
We strengthen their result by showing that the problem remains \NP-hard
if there are only three players and rewards are taken from $\{0,1\}$.

\begin{theorem}\label{thm:posne-np-hard}
\PosNE is \NP-hard, even for turn-based three-player games with
rewards $0$ and~$1$.
\end{theorem}

\begin{proof}
We reduce from the Hamiltonian cycle problem. Given a graph $G=(V,E)$, we
define a turn-based three-player game~$\calG$ as follows: the set of
states is~$V$,  all states are controlled by \pl0, and
the transition function corresponds to~$E$ (\ie $\Gamma_0(v)=vE$
and $\delta(v,\vec{a})=w$ \iff $a_0=w$). Let $n=\abs{V}$ and $v_0\in V$.
\Pl0 receives reward~$1$ in each state. The reward of state~$v_0$ to \pl1
equals~$1$; all other states have reward~$0$ for \pl1.
Finally, \pl2 receives reward $0$ at~$v_0$ and reward~$1$ at all
other states.
\iffull
We show that there is a Hamiltonian cycle in~$G$ \iff $(\calG, v_0)$ has a
positional Nash equilibrium with payoff $\geq(1,1/n,(n-1)/n)$.

($\Rightarrow$) Let $\pi = \pi(0) \pi(1) \dots \pi(n)$ be a Hamiltonian cycle
that starts (and ends) in $\pi(0)=v_0=\pi(n)$. Consider the positional
strategy~$\sigma$ of \pl0 that plays from $\pi(i)$ to $\pi(i+1)$ for all
$i < n$. The induced play from~$v_0$ is the play
$(\pi(0) \pi(1) \ldots \pi(n-1))^\omega$, which
gives payoff~$1$ to \pl0, payoff~$1/n$ to \pl1 and
payoff~$(n-1)/n$ to \pl2. Moreover, it is obvious that we have a
Nash equilibrium.

($\Leftarrow$) Let $\pi$~be the play induced by a positional Nash equilibrium
of $(\calG, v_0)$ with payoff $\geq (1,1/n,(n-1)/n)$.
Since $\pi$~corresponds to a positional strategy profile and gives \pl1
a positive payoff, $\pi$~has the form $\pi=(v_0 v_1 \dots v_{i-1})^\omega$,
where $1\leq i\leq n$ and $v_0 \dots v_{i-1} v_0$ is a simple cycle of~$G$.
Hence, the payoff of~$\pi$ for \pl2 equals $(i-1)/i$. This number is
greater than $(n-1)/n$ only if $i\geq n$. Hence, $i=n$ and
$v_0 \dots v_{i-1} v_0$ is a Hamiltonian cycle.\qed
\fi
\end{proof}

By combining our reduction with a game that has no positional
Nash equilibrium, we can prove the following stronger result for
non-turn-based games.

\begin{corollary}\label{cor:posne-np-complete}
Deciding the existence of a positional Nash equilibrium in a concurrent
limit-average game is \NP-complete, even for three-player games
with rewards $0$ and~$1$.
\end{corollary}

\iffull
\begin{proof}
Membership in \NP follows from \cref{thm:posne-np}. To prove hardness,
we~reduce from the following problem, whose \NP-hardness follows from the
proof of \cref{thm:posne-np-hard}: Given a three-player game $(\calG,s_0)$
with rewards $0$ and~$1$ and $n\in\bbN$ (given in unary), decide whether
$(\calG,s_0)$ has a
positional Nash equilibrium with payoff $\geq (1,1/n,(n-1)/n)$.
From~$\calG$, we construct a new game~$\calG'$, which employs the
game $\calG_2$ from \cref{ex:hide-or-run-2} and is depicted
in \cref{fig:np-reduction}; we set the reward for \pl0 in all states
of~$\calG_2$ to~$1$. Note that we can simulate the fractional rewards in
the terminal state by a cycle of $n$~states with rewards $0$ and~$1$.
\begin{figure}
\centering
\begin{tikzpicture}[x=2.5cm,y=1.4cm,->]
\node (init) at (-0.5,0) {};
\node[play] (0) at (0,0) {$s_0'$};
\node[play] (1) at (1,0) {$s_1'$};
\node[end,label={right:$\begin{map}0 & 0\\ 1 & 1/n\\ 2 & (n-1)/n\end{map}$}] (2) at (2,0) {};
\node (3) at (0,-1) {$(\calG_2,s_1)$};
\node (4) at (1,-1) {$(\calG,s_0)$};

\draw (init) to (0);
\draw (0) to node[above] {$(a,a,a)$} (1);
\draw (0) to node[left] {$(b,a,a)$} (3);
\draw (1) to node[below] {\parbox{1.5cm}{\centering
 $(a,b,a)$\\$(a,a,b)$\\$(a,b,b)$}} (2);
\draw (1) to node[left] {$(a,a,a)$} (4);
\end{tikzpicture}
\caption{\label{fig:np-reduction}The game~$\calG'$}
\end{figure}
We~claim that $(\calG',s_0')$ has a positional Nash equilibrium
\iff $(\calG,s_0)$ has a positional Nash equilibrium with payoff
$\geq (1,1/n,(n-1)/n)$.

$(\Rightarrow)$ Let $\vec{\sigma}$ be a positional Nash equilibrium
of $(\calG',s_0')$. Since $(\calG_2,s_1)$ does not have a Nash equilibrium,
the induced play must either enter the game~$\calG$ or end at the terminal
state with payoff~$0$ for \pl0. But the latter case is impossible since
then \pl0 could improve her payoff by playing action~$b$ at~$s_0'$.
Hence, the induced play enters~$\calG$, and $\vec{\sigma}$~is also
a Nash equilibrium of $(\calG,s_0)$. Moreover, $\vec{\sigma}$~must have
payoff at least $(1,1/n,(n-1)/n)$ since otherwise \pl1 or \pl2 could improve
her payoff by playing action~$b$ at~$s_1'$.

$(\Leftarrow)$ Let $\vec{\sigma}$ be a positional Nash equilibrium
of $(\calG,s_0)$ with payoff at least $(1,1/n,(n-1)/n)$. We can
extend~$\vec{\sigma}$ to a positional Nash equilibrium of $(\calG',s_0')$
by setting $\vec{\sigma}(s_0')=\vec{\sigma}(s_0'(a,a,a)s_1')=(a,a,a)$.\qed
\end{proof}
\fi

\section{Stationary Strategies}
\label{sect:stationary}

To prove the decidability of \StatNE, we appeal to results
established for the \emph{existential theory of the reals}, the set of
all existential first-order sentences
\iffull
(over the appropriate signature)
\fi
that hold in the ordered field $\frakR\coloneq(\bbR,+,\cdot,0,1,\leq)$. The
best known upper bound for the complexity of the associated decision problem
is \PSpace \citep{Canny88}, which leads to the following theorem.

\begin{theorem}\label{thm:statne-pspace}
\StatNE is in \PSpace.
\end{theorem}

\iffull
\begin{proof}
To prove membership in \PSpace, we show that there is a polynomial-time
procedure that on input $\calG,s_0,\vec{x},\vec{y}$ returns an existential
first-order sentence~$\psi$ such that $(\calG,s_0)$ has a stationary
Nash equilibrium with payoff $\geq\vec{x}$ and $\leq\vec{y}$ \iff
$\psi$~holds in~$\frakR$. How does~$\psi$ look like?
Let $\vec{\alpha}=(\alpha^i_{s,a})_{i\in\Pi,s\in S,a\in\Gamma}$,
$\vec{v}=(v^i_s)_{i\in\Pi,s\in S}$,
$\vec{b}=(b_s)_{s\in S}$
and $\vec{z}=(z^i_s)_{i\in\Pi, s\in S}$ be four sets of variables.
The~formula
\[
\phi_i(\vec{\alpha})\coloneq\bigwedge_{s\in S}\Big(
\sum_{a\in\Gamma}\alpha^i_{s,a}=1\wedge
\bigwedge_{\mathmakebox[0.7cm][c]{a\in\Gamma_i(s)}} \alpha^i_{s,a}\geq 0\wedge
\bigwedge_{\mathmakebox[0.7cm][c]{a\in\Gamma\setminus\Gamma_i(s)}}
\alpha^i_{s,a}=0\Big)
\]
states that the mapping~$\sigma_i\colon S\to\bbR^\Gamma$,
defined by $\sigma_i(s)\colon a\mapsto\alpha^i_{s,a}$ is indeed
a stationary strategy for \pli.
Provided that each $\phi_i(\vec{\alpha})$ holds in~$\frakR$, the~formula
\begin{align*}
\eta_i(\vec{\alpha},\vec{z})&\coloneq
\exists\vec{b}\,\Big(\bigwedge_{s\in S}
b_s+z^i_s=r_i(s)+\sum_{\vec{a}\in\Gamma^\Pi}
b_{\delta(s,\vec{a})}\cdot\prod_{j\in\Pi}\alpha^j_{s,a_j}\Big)\:
\wedge \\
&\qquad\bigwedge_{s\in S} z^i_s=\sum_{\vec{a}\in\Gamma^\Pi}
z^i_{\delta(s,\vec{a})}\cdot\prod_{j\in\Pi}\alpha^j_{s,a_j}
\end{align*}
states that $z^i_s=\Exp_s^{\vec{\sigma}}(\pay_i)$ for all $s\in S$,
where $\vec{\sigma}=(\sigma_i)_{i\in\Pi}$
(see \citep[Theorem~8.2.6]{Puterman94}).
Finally, the formula
\begin{align*}
\theta_i(\vec{\alpha},\vec{v})&\coloneq
\exists\vec{b}\,\Big(\bigwedge_{s\in S}\bigwedge_{a\in\Gamma}
b_s+v^i_s\geq r_i(s)+
\sum_{\substack{\vec{a}\in\Gamma^\Pi\\ a_i=a}}
b_{\delta(s,\vec{a})}\cdot
\prod_{j\neq i}\alpha^j_{s,a_j}\Big)\:\wedge \\
&\qquad\bigwedge_{s\in S}\bigwedge_{a\in\Gamma}
v^i_s\geq\sum_{\substack{\vec{a}\in\Gamma^\Pi\\ a_i=a}}
v^i_{\delta(s,\vec{a})}\cdot
\prod_{j\neq i}\alpha^j_{s,a_j}
\end{align*}
states that $\vec{v}$ is a solution of the linear programme for computing
the values of the MDP~$\calG^{\vec{\sigma}_{-i}}$ (see
\citep[Section~9.3]{Puterman94}), \ie the formula
is fulfilled \iff
$v^i_s\geq\sup_{\tau}\Exp^{\vec{\sigma}_{-i},\tau}_s(\pay_i)$
for all $i\in\Pi$ and $s\in S$.

The desired sentence~$\psi$ is the existential closure of the conjunction
of the formulae $\phi_i$, $\eta_i$ and~$\theta_i$ combined
with formulae stating that \pli cannot improve her payoff and that the
expected payoff for \pli lies in-between the given thresholds:
\[\psi\coloneq\exists\vec{\alpha}\,\exists\vec{v}\,\exists\vec{z}\,
\bigwedge_{i\in\Pi}(\phi_i(\vec{\alpha})\wedge\eta_i(\vec{\alpha},\vec{z})
\wedge\theta_i(\vec{\alpha},\vec{v})\wedge v^i_{s_0}\leq z^i_{s_0}\wedge
x_i\leq z^i_{s_0}\leq y_i)\,.\]
Clearly, $\psi$~can be constructed in polynomial time from
$\calG$, $s_0$, $\vec{x}$ and~$\vec{y}$. Moreover, $\psi$~holds
in~$\frakR$ \iff $(\calG,s_0)$ has a stationary Nash equilibrium with
payoff at least $\vec{x}$ and at most~$\vec{y}$.\qed
\end{proof}
\fi

The next theorem shows that \StatNE is \NP-hard, even for turn-based
games with rewards $0$ and~$1$. Note that this does
not follow from the \NP-hardness of \PosNE, but requires a different
proof.

\begin{theorem}\label{thm:statne-np-hard}
\StatNE is NP-hard, even for turn-based games with rewards $0$ and~$1$.
\end{theorem}

\begin{proof}
We employ a reduction from \SAT, which resembles a reduction in
\cite{Ummels08}.
Given a Boolean formula $\phi=C_1\wedge\dots\wedge C_m$
in conjunctive normal form over
propositional variables $X_1,\dots,X_n$, where \wlg $m\geq 1$ and each clause
is nonempty, we build a turn-based game~$\calG$ played by players
$0,1,\dots,n$ as follows: The game~$\calG$ has states
$C_1,\dots,C_m$ controlled by \pl0 and for each clause~$C$ and each
literal~$L$ that occurs in~$C$ a state $(C,L)$, controlled by \pli if
$L=X_i$ or $L=\neg X_i$;
additionally, the~game contains a terminal state~$\bot$. There are transitions
from a clause~$C_j$ to each state~$(C_j,L)$ such that $L$~occurs in~$C_j$ and
from there to~$C_{(j\bmod m)+1}$, and there is a transition
from each state of the form~$(C,\neg X)$ to~$\bot$.
Each state except~$\bot$ has reward~$1$ for \pl0, whereas $\bot$ has reward~$0$
for \pl0. For \pli, all states except states of the form $(C,X_i)$ have
reward~$1$; states of the form $(C,X_i)$ have reward~$0$.
The structure of~$\calG$ is depicted in \cref{fig:sat-reduction}.
\begin{figure*}
\begin{tikzpicture}[x=1.4cm,y=1.3cm,->]
\useasboundingbox (-0.6,3.15) rectangle (6.1,-1.3);
\tikzstyle{dummy}=[circle,minimum size=0.7cm]

\node (c1) at (0,0) [play,label={above left:$0$}] {$C_1$};
\node (l11) at (1,1) [play] {};
\node (l12) at (1,0) {\vdots};
\node (l13) at (1,-1) [play] {};
\node (c2) at (2,0) [play,label={above:$0$}] {$C_2$};
\node (l21) at (3,-1) [dummy] {};
\node (l22) at (3.33,0) [dummy] {\dots};
\node (l23) at (3,1) [dummy] {};
\node (l24) at (3.5,-1) [dummy] {};
\node (l25) at (3.5,1) [dummy] {};
\node (c3) at (4.5,0) [play,label={above:$0$}] {$C_m$};
\node (l31) at (5.5,1) [play] {};
\node (l32) at (5.5,0) [dummy] {\vdots};
\node (l33) at (5.5,-1) [play] {};
\node (bot) at (3.33,2) [end] {$\bot$};

\draw (-0.6,0) to (c1);
\draw (c1) to (l11); \draw (c1) to (l13);
\draw (l11) to (c2); \draw (l13) to (c2);
\draw (c2) to (l21); \draw (c2) to (l23);
\draw (l24) to (c3); \draw (l25) to (c3);
\draw (c3) to (l31); \draw (c3) to (l33);
\draw (l31) .. controls +(110:3.5cm) and +(80:4.5cm) .. (c1);
\draw (l33) .. controls +(65:7.5cm) and +(95:5.5cm) .. (c1);

\draw[dashed] (l11) to [bend left=15] (bot);
\draw[dashed] (l13) to [bend left=20] (bot);
\draw[dashed] (l31) to [bend right=15] (bot);
\draw[dashed] (l33) to [bend right=20] (bot);
\end{tikzpicture}
\caption{\label{fig:sat-reduction}Reducing \SAT to \StatNE}
\end{figure*}
\iffull
\par
Clearly, $\calG$~can be constructed from~$\phi$ in polynomial time.
In~order to establish our reduction, we prove that the
following statements are equivalent:
\begin{enumerate}
\item $\phi$~is satisfiable.
\item $(\calG,C_1)$ has a positional Nash equilibrium with payoff
$\geq 1$ for \pl0.
\item $(\calG,C_1)$ has a stationary Nash equilibrium with payoff
$\geq 1$ for \pl0.
\end{enumerate}

(1.\,$\Rightarrow$\,2.) Assume that $\alpha\colon\{X_1,\dots,X_n\}\to
\{\true,\false\}$ is a satisfying assignment for~$\phi$. We show that the
positional strategy profile~$\vec{\sigma}$ where at any time \pl0 plays from
a clause~$C$ to a fixed state $(C,L)$ such that $L$~is mapped to true
by~$\alpha$ and each \pl{i\neq 0} never plays to~$\bot$ is a Nash
equilibrium of $(\calG,C_1)$ with payoff~$1$ for \pl0.
First note that the induced play never reaches~$\bot$. Hence,
\pl0 receives payoff~$1$, which is the best payoff \pl0 can get.

To show that $\vec{\sigma}$~is a Nash equilibrium, consider any \pl{i\neq 0}
who receives payoff~$<1$. Hence, a state of the form $(C,X_i)$ is visited
in the induced play. However, as \pl0 plays according
to the satisfying assignment, no~state of the form $(C',\neg X_i)$ is ever
visited. Hence, \pli cannot improve her payoff by playing to~$\bot$.

(2.\,$\Rightarrow$\,3.) Trivial.

(3.\,$\Rightarrow$\,1.) Assume that $(\calG,C_1)$ has a stationary
Nash equilibrium~$\vec{\sigma}$ with payoff $\geq 1$ for \pl0.
Hence, the terminal state~$\bot$~is reached with probability~$0$
in~$\vec{\sigma}$. Consider the variable assignment~$\alpha$ that
maps~$X_i$ to true if and only if \pli receives payoff ${<1}$
from~$\vec{\sigma}$; we claim that $\alpha$~satisfies the
formula. Consider any clause~$C$. By the construction of~$\calG$,
there exists a literal $L\in C$ such that $\sigma_0((C,L)\mid C)>0$.
If $L=X_i$, then $\Exp_{C_1}^{\vec{\sigma}}(\pay_i)<1$ and
$\alpha$~maps~$X_i$ to true, thus satisfying~$C$. If $L=\neg X_i$,
then \pli must receive payoff~$1$ since otherwise she could
switch to the positional strategy~$\tau$ that plays from~$(C,L)$
to~$\bot$; in the strategy profile $(\vec{\sigma}_{-i},\tau)$
the state~$\bot$ is visited with probability~$1$, which gives
payoff~$1$ to \pli. Hence, $\alpha$ maps~$X_i$ to false and
satisfies~$C$.\qed
\fi
\end{proof}

By combining our reduction with the game from \cref{ex:hide-or-run}, we can
prove the following stronger result for concurrent games.

\begin{corollary}\label{cor:statnene-np-complete}
Deciding the existence of a stationary Nash equilibrium in a concurrent
limit-average game with rewards $0$ and~$1$ is \NP-hard.
\end{corollary}

\iffull
\begin{proof}
The proof is similar to the proof of \cref{cor:posne-np-complete}.
From a given concurrent limit-average game $(\calG,s_0)$ with
rewards $0$ and~$1$, we construct a new game $(\calG',s_0')$ such
that $(\calG',s_0')$ has a stationary Nash equilibrium \iff
$(\calG,s_0)$ has a stationary Nash equilibrium with payoff
at least~$1$ for \pl0.
The game~$\calG'$ is the disjoint union of~$\calG$,
the game $\calG_2$ from \cref{ex:hide-or-run-2}, and the state~$s_0'$,
which is controlled by \pl0. At~$s_0'$ \pl0 can
either play to the initial state~$s_0$ of~$\calG$ or to the
initial state~$s_1$ of~$\calG_2$. Finally, we set the reward for
\pl0 in every state of~$\calG_2$ to~$1$.\qed
\end{proof}
\fi

So far we have shown that \StatNE is contained in \PSpace and hard
for \NP, leaving a considerable gap between the two bounds.
In order to gain a better
understanding of \StatNE, we relate this problem to the
\emph{square root sum problem} (\SqrtSum), an important problem about
numerical computations. Formally, \SqrtSum is the following decision
problem: Given numbers $d_1,\dots,d_n,k\in\bbN$, decide whether
$\sum_{i=1}^n \sqrt{d_i}\geq k$.
Recently, \citet{AllenderBKM09} showed that \SqrtSum belongs to the fourth
level of the \emph{counting hierarchy}, a slight improvement
over the previously known \PSpace upper bound.
However, it has been an open question since the 1970s as to whether \SqrtSum
falls into the polynomial hierarchy \citep{GareyGJ76,EtessamiY10}. We give
a polynomial-time reduction from \SqrtSum to \StatNE for turn-based
terminal-reward games.
Hence, \StatNE is at least as hard as \SqrtSum, and showing that
\StatNE resides inside the polynomial hierarchy would imply a major
breakthrough in understanding the complexity of numerical computations.
While our reduction is similar to the one in \citep{UmmelsW09}, it
requires new techniques to simulate stochastic states.

\begin{theorem}\label{thm:statne-sqrtsum}
\SqrtSum is polynomial-time reducible to \StatNE for turn-based
8-player terminal-reward games.
\end{theorem}

\iffull
Before we state the reduction, let us first examine the
game $\calG(p)$, where $0\leq p\leq 1$, which is played
by players $0,1,\dots,5$ and depicted in \cref{fig:sqrt-reduction-1}.
\begin{figure}
\begin{tikzpicture}[x=1.6cm,y=1.3cm,->]
\node (init) at (-0.7,0) {};
\node[play,label={above:$2$}] (1) at (0,0) {$s_1$};
\node[end,label={right:$\begin{map}0 & -1\\ 2 & \phantom{+}1\end{map}$}] (1a) at (1,0) {};
\node[play,label={-85:$4$}] (11) at (0.5,1.1) {$r_1$};
\node[end,label={right:$\begin{map}0 & -1\\ 4 & \phantom{+}1\end{map}$}] (11a) at (0.5,2.1) {};
\node[play,label={below:$5$}] (12) at (1.7,1.1) {$t_1$};
\node[end,label={right:$\begin{map}0 & \phantom{1}{-1}\\ 5 & 1{-}p\end{map}$}] (12a) at (1.7,2.1) {};
\node[play,label={below:$0$}] (2) at (3,1.1) {$u_1$};
\node[end,label={right:$\left\{\begin{map}1 & 1\\ 2 & 1\\ 4 & 2{-}p\end{map}\right.$}] (2a) at (3,2.1) {};
\node[play,label={-95:$0$}] (3) at (4.5,1.1) {$v_1$};
\node[end,label={right:$\left\{\begin{map}2 & 2\\ 4 & 1{-}p\\ 5 & 1\end{map}\right.$}] (3a) at (4.5,2.1) {};
\node[play,label={right:$3$}] (4) at (5,0) {$s_2$};
\node[end,label={left:$\begin{map}0 & -1\\ 3 & \phantom{+}1\end{map}$}] (4a) at (4,0) {};
\node[play,label={95:$4$}] (13) at (4.5,-1.1) {$r_2$};
\node[end,label={right:$\begin{map}0 & \phantom{1}{-1}\\ 4 & 1{-}p\end{map}$}] (13a) at (4.5,-2.1) {};
\node[play,label={above:$5$}] (14) at (3.3,-1.1) {$t_2$};
\node[end,label={right:$\begin{map}0 & -1\\ 5 & \phantom{+}1\end{map}$}] (14a) at (3.3,-2.1) {};
\node[play,label={above:$0$}] (5) at (1.9,-1.1) {$u_2$};
\node[end,label={right:$\left\{\begin{map}1 & 1\\ 3 & 1\\ 5 & 2{-}p\end{map}\right.$}] (5a) at (1.9,-2.1) {};
\node[play,label={85:$0$}] (6) at (0.5,-1.1) {$v_2$};
\node[end,label={right:$\left\{\begin{map}3 & 2\\ 4 & 1\\ 5 & 1{-}p\end{map}\right.$}] (6a) at (0.5,-2.1) {};

\draw (init) to (1);
\draw (1) to (11); \draw (1) to (1a);
\draw (11) to (12); \draw (11) to (11a);
\draw (12) to (2); \draw (12) to (12a);
\draw (2) to (3); \draw (2) to (2a);
\draw (3) to (4); \draw (3) to (3a);
\draw (4) to (13); \draw (4) to (4a);
\draw (13) to (14); \draw (13) to (13a);
\draw (14) to (5); \draw (14) to (14a);
\draw (5) to (6); \draw (5) to (5a);
\draw (6) to (1); \draw (6) to (6a);
\end{tikzpicture}
\caption{\label{fig:sqrt-reduction-1}The game~$\calG(p)$}
\end{figure}

\begin{lemma}\label{lemma:statne-reduction}
The maximal payoff \pl1 receives in a stationary Nash equilibrium of
$(\calG(p),s_1)$ where \pl0 receives payoff $\geq 0$ equals $\sqrt{p}$.
\end{lemma}

\begin{proof}
Let $\vec{\sigma}$ be a stationary strategy profile of $(\calG(p),s_1)$
where \pl0 receives payoff $\geq 0$, and let $q_i=\sigma_0(v_i\mid u_i)$
be the probability that \pl0 moves from $u_i$ to~$v_i$. We claim that
$q\coloneq q_1=q_2=1-p$ if $\vec{\sigma}$~is a Nash equilibrium. Let
$z=\Exp^{\vec{\sigma}}_{v_1}(\pay_4)$ and
$z'=\Exp^{\vec{\sigma}}_{v_1}(\pay_5)$. Since $\vec{\sigma}$~is
a Nash equilibrium, we have $z\geq 1-p$ and $z'\geq 1$ (otherwise
\pl4 or \pl5 would prefer to leave the game at $r_2$ or~$t_2$).
On the other hand, since at every terminal state the sum of the rewards
for players $4$ and~$5$ is at most $2-p$, we have $z+z'\leq 2-p$.
Hence, $z=1-p$ and $z'=1$. Now
consider the expected payoffs for players $4$ and~$5$ from~$r_1$:
\begin{align*}
\Exp^{\vec{\sigma}}_{r_1}(\pay_4)
&=(1-q_1)(2-p)+q_1\cdot z=2-q_1-p; \\
\Exp^{\vec{\sigma}}_{r_1}(\pay_5)
&= q_1\cdot z'= q_1\,.
\end{align*}
Since $\vec{\sigma}$~is a Nash equilibrium, these numbers are bounded
from below by $1$ and~$1-p$, respectively (otherwise, \pl4 or \pl5 would
leave the game at $r_1$ or~$t_1$). Hence, $q_1=1-p$. The reasoning that
$q_2=1-p$ is analogous.

In the following, assume without loss of generality that $0<p<1$
(otherwise the statement of the lemma is trivial). For any stationary strategy
profile~$\vec{\sigma}$ of~$\calG(p)$ where \pl0 receives payoff
$\geq 0$, let $x_1=\sigma_0(s_2\mid v_1)$ and $x_2=\sigma_0(s_1\mid v_2)$ be
the probabilities that \pl0 does not leave the game at~$v_1$,
respectively~$v_2$. Given $x_1$ and~$x_2$, for $i=1,2$ we can compute
the payoff $f_i(x_1,x_2)\coloneq\Exp^{\vec{\sigma}}_{s_i}(\pay_{i+1})$
for \pl{i+1} from~$s_i$ by
\[f_i(x_1,x_2)=\frac{p+2q(1-x_i)}{1-q^2x_1x_2}\,.\]
To have a Nash equilibrium, it must be the case that
$f_1(x_1,x_2),f_2(x_1,x_2)\geq 1$ since otherwise \pl2 or \pl3 would prefer
to leave the game at $s_1$ or~$s_2$, respectively, which would give the
respective player payoff~$1$ immediately.
Vice versa, if $f_1(x_1,x_2),f_2(x_1,x_2)\geq 1$ then $\vec{\sigma}$
is a Nash equilibrium with expected payoff
\[f(x_1,x_2)\coloneq\frac{p+q x_1p}{1-q^2x_1x_2}\]
for \pl1. Hence,
to determine the maximum payoff for \pl1 in a stationary Nash equilibrium
where \pl0 receives payoff $\geq 0$, we have to maximise
$f(x_1,x_2)$ under the constraints $f_1(x_1,x_2),f_2(x_1,x_2)\geq 1$ and
$0\leq x_1,x_2\leq 1$. We claim that the maximum is reached only if
$x_1=x_2$. If \eg $x_1>x_2$, then we can achieve a higher payoff for \pl1
by setting $x_2'\coloneq x_1$, and the constraints are still satisfied:
\[\frac{p+2q(1-x_2')}{1-q^2x_1x_2'}=\frac{p+2q(1-x_1)}{1-q^2 x_1^2}
 \geq\frac{p+2q(1-x_1)}{1-q^2 x_1 x_2}\geq 1\,.\]
Hence, it suffices to maximise $f(x,x)$ subject to
$f_1(x,x)\geq 1$ and ${0\leq x\leq 1}$, which is
equivalent to maximising $f(x,x)$ subject to $(1-p)x^2-2x+1\geq 0$
and $0\leq x\leq 1$. and The roots of the quadratic function are
$(1\pm\sqrt{p})/(1-p)$, but $(1+\sqrt{p})/(1-p)>1$ for $p>0$.
Therefore, any solution~$x$
must satisfy $x\leq x_0\coloneq (1-\sqrt{p})/(1-p)$. Since
$0\leq x_0\leq 1$ for $0<p<1$ and $f(x,x)$ is strictly increasing on
$[0,1]$, the optimal solution is~$x_0$, and the maximal payoff for \pl1
in a stationary Nash equilibrium of $(\calG(p),s_1)$ where \pl0
receives payoff $\geq 0$ equals indeed
\[f(x_0,x_0)
=\frac{p+qx_0 p}{1-q^2x_0^2}
=\frac{p}{1-q x_0}
=\frac{p}{1-(1-p)x_0}
=\frac{p}{1-(1-\sqrt{p})}=\sqrt{p}\,.\tag*{\qedsymbol}\]
\end{proof}

\begin{proof}[of \cref{thm:statne-sqrtsum}]
Given an instance $(d_1,\dots,d_n,k)$ of \SqrtSum,
where \wlg $n>0$, $d_i>0$ for each $i=1,\dots,n$, and
$d\coloneq\sum_{i=1}^n d_i$,
we construct a turn-based 8-player terminal-reward game $(\calG,s)$
such that $(\calG,s)$ has a stationary Nash equilibrium with payoff
$\geq (0,\frac{k}{d(n+1)},0,\dots,0)$ \iff $\sum_{i=1}^n\sqrt{d_i}\geq k$.
Define $p_i\coloneq d_i/d^2$ for $i=1,\dots,n$.
For the reduction, we use $n$~copies of the game~$\calG(p)$, where in the
$i$th copy we set $p$ to~$p_i$; in each copy, we set the rewards to \pl6 and
\pl7 at all terminal states to $1$ and~$0$, respectively. The complete
game~$\calG$ is depicted in \cref{fig:sqrt-reduction-2};
\begin{figure}
\begin{tikzpicture}[x=1.5cm,y=1.3cm,->]
\node (init) at (-0.6,0) {};
\node (0) at (0,0) [play,label={above:$6$}] {$s_n$};
\node (0a) at (0,-1) [end,label={below:$\begin{map}0 & -1\\ 6 & \frac{n}{n+1}\end{map}$}] {};
\node (1) at (1,0) [play,label={above:$7$}] {$r_n$};
\node (1a) at (1,-1) [end,label={below:$\begin{map}0 & -1\\ 7 & \frac{n+1}{n+1}\end{map}$}] {};
\node (2) at (2,0) [play,label={above:$0$}] {$t_n$};
\node (2a) at (2,-1) {$\calG(p_n)$};

\node (23) at (2.75,0) {$\cdots$};

\node (3) at (3.5,0) [play,label={above:$6$}] {$s_i$};
\node (3a) at (3.5,-1) [end,label={below:$\begin{map}0 & -1\\ 6 & \frac{i}{i+1}\end{map}$}] {};
\node (4) at (4.5,0) [play,label={above:$7$}] {$r_i$};
\node (4a) at (4.5,-1) [end,label={below:$\begin{map}0 & -1\\ 7 & \frac{n+1}{i+1}\end{map}$}] {};
\node (5) at (5.5,0) [play,label={above:$0$}] {$t_i$};
\node (5a) at (5.5,-1) {$\calG(p_i)$};

\node (56) at (6.25,0) {$\cdots$};

\node (6) at (7,0) [end,label={above:$7\colon n{+}1$}] {$s_0$};

\draw (init) to (0);
\draw (0) to (0a); \draw (0) to (1);
\draw (1) to (1a); \draw (1) to (2);
\draw (2) to (2a); \draw (2) to (23);
\draw (23) to (3);
\draw (3) to (3a); \draw (3) to (4);
\draw (4) to (4a); \draw (4) to (5);
\draw (5) to (5a); \draw (5) to (56);
\draw (56) to (6);
\end{tikzpicture}
\caption{\label{fig:sqrt-reduction-2}Reducing \SqrtSum to \StatNE}
\end{figure}
it can obviously be constructed in polynomial time. We~claim
that in any (stationary) Nash equilibrium of $(\calG,s_n)$ where \pl0
receives
payoff~$\geq 0$ the probability of reaching the game~$\calG(p_i)$
equals~$1/(n+1)$ for all $i=1,\dots,n$.
First note that in any such
equilibrium the state~$s_0$ must be reached with
positive probability since otherwise \pl7 would prefer to leave the game
at one of the states~$r_i$, giving \pl0 payoff~$<0$.
Now let $\vec{\sigma}$ be a stationary Nash equilibrium of $(\calG,s_n)$
where \pl0 receives payoff $\geq 0$, and set
$q_i\coloneq\sigma_0(s_{i-1}\mid t_i)$.
By induction on~$i$, we prove that
$q_i=i/(i+1)$. For $i=1$, this is true because if $q_i>\frac{1}{2}$
then \pl6 would prefer to leave the game at~$s_1$, and if
$q_i<\frac{1}{2}$ then \pl7 would prefer to leave the game at~$r_1$.
Now let $i>1$ and assume that $q_j=j/(j+1)$ for all $j<i$. A simple
calculation reveals that the expected payoffs for \pl6 and \pl7
from~$s_{i-1}$ equal $(i-1)/i$ and $(n+1)/i$, respectively.
Hence, the expected payoff for \pl6 from state~$t_i$ equals
\[1-q_i+q_i\cdot\frac{i-1}{i}=1-\frac{q_i}{i}
=\frac{i+1-q_i\cdot\frac{i+1}{i}}{i+1}\,.\]
If $q_i>i/(i+1)$, then this number would be strictly smaller than
$i/(i+1)$, and \pl6 would be better off by leaving the game at~$s_i$.
On the other hand, the expected payoff for \pl7 from state~$t_i$ equals
$q_i(n+1)/i$. If $q_i<i/(i+1)$, then this number would
be strictly smaller than $(n+1)/(i+1)$, and \pl7 would prefer to
leave the game at~$r_i$. In both cases, we have a contradiction
to $\vec{\sigma}$ being a Nash equilibrium. Hence, $q_i=i/(i+1)$
for all $i=1,\ldots,n$, and
the probability of reaching the game~$\calG(p_i)$ from~$s_n$ equals
\[(1-q_i)\prod_{j=i+1}^n q_i=(1-\frac{i}{i+1})\prod_{j=i+1}^n\frac{j}{j+1}=\frac{1}{i+1}
\cdot \frac{i+1}{n+1} = \frac{1}{n+1}\,.\]
It remains to be shown that $(\calG,s_n)$ has a stationary Nash
equilibrium with payoff $\geq(0,\frac{k}{d(n+1)},0,\dots,0)$ \iff
$\sum_{i=1}^n\sqrt{d_i}\geq k$. By \cref{lemma:statne-reduction},
the maximal payoff \pl1 receives in a stationary Nash equilibrium of $(\calG(p_i),s_1)$ where \pl0 receives payoff at least~$0$ equals
$\sqrt{p_i}=\sqrt{d_i}/d$. Hence,
the maximal payoff \pl1 receives in a stationary Nash equilibrium
of $(\calG,s_n)$ where \pl0 receives payoff at least~$0$ equals
\[\sum_{i=1}^n\frac{1}{n+1}\cdot\frac{\sqrt{d_i}}{d}
=\frac{1}{d(n+1)}\cdot\sum_{i=1}^n\sqrt{d_i}\,.\]
We conclude that $(\calG,s_n)$ has a stationary Nash equilibrium with
payoff $\geq(0,\frac{k}{d(n+1)},0,\dots,0)$ \iff
$\sum_{i=1}^n\sqrt{d_i}\geq k$.\qed
\end{proof}

\fi

Again, we can combine our reduction with the game from \cref{ex:hide-or-run}
to prove a stronger result for games that are not turn-based.

\begin{corollary}
Deciding whether a concurrent 8-player terminal reward game has a stationary
Nash equilibrium is hard for \SqrtSum.
\end{corollary}

\iffull
\begin{proof}
The proof is analogous to the proof of \cref{cor:posne-np-complete}, but
we use the game~$\calG_1$ from \cref{ex:hide-or-run} instead of the
game~$\calG_2$, and
\pl0 receives reward~$0$ in each state of $\calG_1$ and reward~$-1$ in
the new terminal state. Since $\calG_1$~is a terminal-reward game, the
resulting game~$\calG'$ is a terminal-reward
game if the original game~$\calG$ is a terminal-reward game.\qed
\end{proof}
\fi

\begin{remark}
\iffull
The positive results of \cref{sect:positional,sect:stationary} can easily
be extended to equilibria in pure or randomised strategies with a memory of a
fixed size $k\in\bbN$: a~nondeterministic algorithm can guess a
memory structure~$\frakM$ of size~$k$
and then look for a positional, respectively stationary, equilibrium
in the product of the original game~$\calG$ with the memory~$\frakM$.
Hence, for any fixed $k\in\bbN$, we~can decide in \PSpace (\NP) the existence
of a randomised (pure) equilibrium of size~$k$ with payoff $\geq\vec{x}$ and
$\leq\vec{y}$. Moreover, these results extend to stochastic games
(by~appealing to results on MDPs with limit-average objectives;
see \eg~\cite{Puterman94}).
\fi
\end{remark}

\section{Pure Strategies}
\label{sect:pure}

In this section, we show that \PureNE is decidable and, in fact, \NP-complete.
Let $\calG$ be a concurrent game, $s\in S$ and $i\in\Pi$. We define
\[\pval^\calG_i(s)=\inf\nolimits_{\vec{\sigma}}\sup\nolimits_{\tau} \Exp^{\vec{\sigma}_{-i},\tau}_{s}(\pay_i),\]
where $\vec{\sigma}$~ranges over all \emph{pure} strategy profiles of~$\calG$
and $\tau$~ranges over all strategies of \pli. Intuitively,
$\pval^{\calG}_i(s)$ is the lowest payoff that the coalition
$\Pi\setminus\{i\}$ can inflict on \pli by playing a pure strategy.

By a reduction to a turn-based two-player zero-sum game, we can show
that there is a positional strategy profile that attains this value.

\begin{proposition}\label{prop:pval}
Let $\calG$ be a concurrent game, and $i\in\Pi$.
There exists a positional strategy profile~$\vec{\sigma}^*$ such that
$\Exp^{\vec{\sigma}^*_{-i},\tau}_{s}(\pay_i)\leq\pval^{\calG}_i(s)$
for all states~$s$ and all strategies~$\tau$ of \pli.
\end{proposition}

\iffull
\begin{proof}
We define a turn-based two-player zero-sum game $\calG'$ with players $0$
and~$1$ as follows:
The set of states of~$\calG'$ is
$S'=S\cup(S\times\Gamma^\Pi)$. At a state $s\in S$, \pl1
chooses an action profile~$\vec{a}$ that is legal at~$s$,
which leads the game to the state $(s,\vec{a})$. At a state of the form
$(s,\vec{a})$, \pl0 chooses an action $b\in\Gamma_i(s)$, which leads the
game to the state $\delta(s,(\vec{a}_{-i},b))$. Finally, \pl0's reward at
a state $s\in S$ or $(s,\vec{a})\in S\times\Gamma^\Pi$ is
$r'(s)=r'(s,\vec{a})=r_i(s)$ (and \pl1's reward is the opposite).
By \cite{EhrenfeuchtM79}, there exists a function $\nu\colon S'\to\bbQ$
(the \emph{value} function) and positional strategies $\sigma^*$ and~$\tau^*$
for \pl1 and \pl0, respectively, such that
$\Exp_{s}^{\tau,\sigma^*}(\pay'_0)\leq \nu(s)$ for all $s\in S'$
and all strategies~$\tau$ of \pl0 in~$\calG'$,
and $\Exp_{s}^{\tau^*,\sigma}(\pay'_0)\geq \nu(s)$ for all $s\in S'$ and all
strategies~$\sigma$ of \pl1 in~$\calG'$.
We can translate \pl1's strategy~$\sigma^*$
into a positional strategy profile~$\vec{\sigma}^*$ of~$\calG$ such that
$\Exp_{s}^{\vec{\sigma}^*_{-i},\tau}(\pay_i)\leq \nu(s)$ for all states
$s\in S$ and all strategies~$\tau$ of \pli in~$\calG$. Hence,
$\pval^{\calG}_i(s)\leq\sup_\tau\Exp_{s}^{\vec{\sigma}^*_{-i},\tau}(\pay_i)
\leq \nu(s)$ for all $s\in S$.
We claim that $\pval^{\calG}_i(s)\geq\nu(s)$ for all $s\in S$, which implies
that $\pval^{\calG}_i(s)=\nu(s)$ for all $s\in S$ and that $\vec{\sigma}^*$~is
the strategy profile we are looking for. Otherwise, there would exist a pure
strategy profile~$\vec{\sigma}$ in~$\calG$ such that
$\sup_\tau\Exp_{s}^{\vec{\sigma}_{-i},\tau}(\pay_i)<\nu(s)$ for some
$s\in S$.
But we could translate such a strategy profile~$\vec{\sigma}$ into a
pure strategy~$\sigma$ of \pl1 in~$\calG'$ such that
$\Exp_s^{\tau^*\!,\sigma}(\pay'_0)<\nu(s)$, a contradiction to the optimality
of~$\tau^*$.\qed
\end{proof}
\fi

Given a payoff vector $\vec{z}\in(\bbR\cup\{\pm\infty\})^\Pi$, we define
a directed graph $G(\vec{z})=(V,E)$ (with self-loops) as
follows: $V=S$, and there is an edge from $s$ to~$t$ \iff there is an
action profile~$\vec{a}$ with $\delta(s,\vec{a})=t$ such that
(1)~$\vec{a}$~is legal at~$s$ and
(2)~$\pval^{\calG}_i(\delta(s,(\vec{a}_{-i},b)))\leq z_i$
for each \pli and each action $b\in\Gamma_i(s)$. Following
\cite{BouyerBM10b}, we~call any~$\vec{a}$ that fulfils (1) and~(2)
\emph{$\vec{z}$-secure} at~$s$.

\begin{lemma}\label{lemma:secure-1}
Let $\vec{z}\in(\bbR\cup\{\pm\infty\})^\Pi\!$.
If there exists an infinite path~$\pi$ in~$G(\vec{z})$ from~$s_0$ with
$z_i\leq\pay_i(\pi)$ for each \pli, then $(\calG,s_0)$ has a pure Nash
equilibrium with payoff~$\pay_i(\pi)$ for \pli.
\end{lemma}

\begin{proof}
Let $\pi=s_0s_1\dots$ be an infinite path in~$G(\vec{z})$ from~$s_0$ with
$z_i\leq\pay_i(\pi)$ for each \pli. We define a pure strategy
profile~$\vec{\sigma}$ as follows: For histories of the form
$x=s_0\vec{a}_0s_1\dots s_{k-1}\vec{a}_{k-1}s_k$, we set
$\vec{\sigma}(x)$ to an action profile~$\vec{a}$ with
$\delta(s_k,\vec{a})=s_{k+1}$ that is $\vec{z}$-secure at~$s_k$.
For all other histories~$x=t_0 \vec{a}_0 t_1\dots t_{k-1}\vec{a}_{k-1}t_k$,
consider the least~$j$ such that $s_{j+1}\neq t_{j+1}$.
If $\vec{a}_j$~differs from a $\vec{z}$-secure action
profile~$\vec{a}$ at~$s_j$ in
precisely one entry~$i$, we set $\vec{\sigma}(x)=\vec{\sigma}^*(t_k)$,
where $\vec{\sigma}^*$~is a (fixed) positional strategy profile such that
$\Exp^{\vec{\sigma}^*_{-i},\tau}_{s}(\pay_i)\leq\pval^{\calG}_i(s)$
for all $s\in S$
(which is guaranteed to exist by \cref{prop:pval}); otherwise,
$\vec{\sigma}(x)$~can be chosen arbitrarily. It is easy to see that
$\vec{\sigma}$~is a Nash equilibrium with induced play~$\pi$.\qed
\end{proof}

\begin{lemma}\label{lemma:secure-2}
Let $\vec{\sigma}$ be a pure Nash equilibrium of $(\calG,s_0)$ with
payoff~$\vec{z}$. Then there exists an infinite path~$\pi$ in~$G(\vec{z})$
from~$s_0$ with $\pay_i(\pi)=z_i$ for each \pli.
\end{lemma}

\begin{proof}
Let $s_0\vec{a}_0s_1\vec{a}_1\ldots$ be the play induced by~$\vec{\sigma}$.
We claim that $\pi\coloneq s_0 s_1\ldots$~is a path in~$G(\vec{z})$.
Otherwise, consider the least~$k$ such that $(s_k,s_{k+1})$
is not an edge in~$G(\vec{z})$. Hence, there exists no $\vec{z}$-secure
action profile at $s\coloneq s_k$. Since $\vec{a}_k$~is certainly legal
at~$s$, there exists a \pli and an action $b\in\Gamma_i(s)$ such that
$\pval^{\calG}_i(\delta(s,(\vec{a}_{-i},b)))>z_i$. But then \pli can
improve her payoff by switching to a strategy that mimics~$\sigma_i$
until $s$~is reached, then plays action~$b$, and after that
mimics a strategy that ensures payoff~$>z_i$ against any pure
strategy profile.
This contradicts the assumption that $\vec{\sigma}$~is a Nash equilibrium.\qed
\end{proof}

Using \cref{lemma:secure-1,lemma:secure-2}, we can reduce the task of
finding a pure Nash equilibrium to the task of finding a path in a
multi-weighted graph whose limit-average weight vector falls between two
thresholds.
The latter problem can be solved in polynomial time
by solving a linear programme with one variable for each pair of
\iffull
a weight function and an edge in the graph, as we prove in the appendix.
\fi

\begin{theorem}\label{thm:mp-path}
Given a finite directed graph $G=(V,E)$ with weight functions
$r_0,\ldots,r_{k-1}\colon V\to\bbQ$, $v_0\in V$, and
$\vec{x},\vec{y}\in(\bbQ\cup\{\pm\infty\})^k$, we can decide in polynomial
time whether there exists an infinite path~$\pi=v_0 v_1\ldots$ in~$G$ with
$x_i\leq\liminf_{n\to\infty}\frac{1}{n}\sum_{j=0}^{n-1}r_i(v_j)\leq y_i$
for all $i=0,\ldots,k-1$.
\end{theorem}

We can now describe a nondeterministic algorithm to decide the existence
of a pure Nash equilibrium with payoff $\geq\vec{x}$ and $\leq\vec{y}$ in
polynomial time. The~algorithm starts by guessing, for each \pli, a
positional strategy profile~$\vec{\sigma}^i$ of~$\calG$ and
computes $p_i(s)\coloneq\sup_\tau\Exp^{\vec{\sigma}_{-i}^i,\tau}_{s}(\pay_i)$
for each $s\in S$; these numbers can be computed in polynomial time using the
algorithm given by \citet{Karp78}.
The algorithm then guesses a vector~$\vec{z}\in(\bbR\cup\{\pm\infty\})^\Pi$
by setting~$z_i$
either to~$x_i$ or to~$p_i(s)$ for some $s\in S$ with $x_i\leq p_i(s)$, and
constructs the graph~$G'(\vec{z})$, which is defined as $G(\vec{z})$ but
with $p_i(s)$ substituted for $\pval^{\calG}_i(s)$.
Finally, the algorithm determines (in polynomial time) whether there exists an
infinite path~$\pi$ in~$G(\vec{z})$ from~$s_0$ with
$z_i\leq\pay_i(\pi)\leq y_i$ for all $i\in\Pi$. If such a path exists, the
algorithm accepts; otherwise it rejects.

\begin{theorem}\label{thm:purene-np}
\PureNE is in \NP.
\end{theorem}

\begin{proof}
We claim that the algorithm described above is correct, \ie sound
and complete.
To prove soundness, assume that the algorithm accepts its input. Hence,
there exists an infinite path~$\pi$ in $G'(\vec{z})$ from~$s_0$
with $z_i\leq\pay_i(\pi)\leq y_i$. Since
$\pval^{\calG}_i(s)\leq p_i(s)$ for all $i\in\Pi$ and $s\in S$, the
graph $G'(\vec{z})$ is a subgraph of~$G(\vec{z})$. Hence,
$\pi$~is also an infinite path in~$G(\vec{z})$. By
\cref{lemma:secure-1}, we can conclude that $(\calG,s_0)$ has a pure Nash
equilibrium with payoff $\geq\vec{z}\geq{\vec{x}}$ and $\leq\vec{y}$.

To prove that the algorithm is complete, let $\vec{\sigma}$ be a pure
Nash equilibrium of $(\calG,s_0)$ with payoff~$\vec{z}$, where
$\vec{x}\leq\vec{z}\leq\vec{y}$. By \cref{prop:pval}, the algorithm can
guess positional strategy profiles~$\vec{\sigma}^i$ such that
$p_i(s)=\pval^{\calG}_i(s)$ for all ${s\in S}$. If the algorithm
additionally guesses the payoff vector~$\vec{z}'$ defined by
$z_i'=\max\{x_i,\pval^{\calG}_i(s):s\in S,\pval^{\calG}_i(s)\leq z_i\}$
for all $i\in\Pi$,
then the graph~$G(\vec{z})$ coincides with the graph~$G(\vec{z}')$ (and
thus with $G'(\vec{z}')$).
By \cref{lemma:secure-2}, there exists an infinite path~$\pi$ in~$G(\vec{z})$
from~$s_0$ such that $z_i'\leq z_i=\pay_i(\pi)\leq y_i$ for all $i\in\Pi$.
Hence, the algorithm accepts.\qed
\end{proof}

The following theorem shows that \PureNE is \NP-hard. In fact, \NP-hardness
holds even for turn-based games with rewards $0$ and~$1$.

\begin{theorem}\label{thm:purene-np-hard}
\PureNE is \NP-hard, even for turn-based games with rewards $0$~and~$1$.
\end{theorem}

\begin{proof}
Again, we reduce from \SAT.
Given a Boolean formula $\phi=C_1\wedge\dots\wedge C_m$ in conjunctive normal
form over propositional variables $X_1,\dots,X_n$, where \wlg $m\geq 1$ and
each clause is nonempty, let $\calG$ be the turn-based game described in
the proof of \cref{thm:statne-np-hard} and depicted in
\iffull
\cref{fig:sat-reduction}.
We claim that the following statements are equivalent:
\begin{enumerate}
\item $\phi$~is satisfiable.
\item $(\calG,C_1)$ has a positional Nash equilibrium with payoff
$\geq 1$ for \pl0.
\item $(\calG,C_1)$ has a pure Nash equilibrium with payoff
$\geq 1$ for \pl0.
\end{enumerate}

Since the implication (1.\,$\Rightarrow$\,2.) was already proved in the
proof of \cref{thm:statne-np-hard} and the implication (2.\,$\Rightarrow$\,3.)
is trivial, we only need to prove that 3.\ implies 1.
Hence, assume that $(\calG,C_1)$ has a pure Nash
equilibrium~$\vec{\sigma}$ with payoff $\geq 1$ for \pl0.
Since \pl0 receives payoff $\geq 1$, the terminal state~$\bot$~is not
reached in the induced play~$\pi$. Consider the variable
assignment~$\alpha$ that maps~$X_i$ to true if and only if \pli
receives payoff ${<1}$ from~$\pi$; we claim that $\alpha$~satisfies the
formula. Consider any clause~$C$.
Set $T=\{(C,X_i),(C,\neg X_i):i=1,\ldots,n\}$, and denote by $\One_s$
the characteristic function of $s\in T$. We have
\[
\sum_{s\in T}\liminf_{n\to\infty}
 \frac{1}{n}\sum_{j=0}^{n-1}-\One_s(\pi(j))
\leq
 \liminf_{n\to\infty}
 \frac{1}{n}\sum_{j=0}^{n-1}\sum_{s\in T}-\One_s(\pi(j))
=-\frac{1}{2m}<0\,.\]
In particular, there exists a state $s=(C,L)$ such that
\[\liminf_{n\to\infty}\frac{1}{n}\sum_{j=0}^{n-1}-\One_s(\pi(j))<0\,.\]
If $L=X_i$, then $r_i\leq 1-\One_s$. Hence, $\pay_i(\pi)<1$, and
$\alpha$~maps~$X_i$ to true, thereby satisfying~$C$.
If $L=\neg X_i$, then \pli must receive payoff~$1$, because otherwise
she could improve her payoff by playing from $s$ to~$\bot$. Hence,
$\alpha$~maps~$X_i$ to false and satisfies~$C$.\qed
\fi
\end{proof}

It follows from \cref{thm:purene-np,thm:purene-np-hard} that \PureNE
is \NP-complete. By combining our reduction with a game that has no
pure Nash equilibrium, we can prove the following stronger result for
non-turn-based games.

\begin{corollary}\label{cor:purene-np-complete}
Deciding the existence of a pure Nash equilibrium in a concurrent
limit-average game is \NP-complete, even for games with rewards
$0$ and~$1$.
\end{corollary}

\iffull
\begin{proof}
The proof is analogous to the proof of \cref{cor:statnene-np-complete}.\qed
\end{proof}
\fi

Note that \cref{thm:purene-np-hard,cor:purene-np-complete} do not apply to
terminal-reward games. In fact, \PureNE is decidable in \PTime for these
games, which follows from two facts about terminal-reward games: (1)~the
numbers $\pval_i^{\calG}(s)$ can be computed in polynomial time
(using a reduction to a turn-based two-player zero-sum game and
applying a result of \citet{Washburn90}), and (2)~the only possible
vectors that can emerge as the payoff of a pure strategy profile
are the zero vector and the reward vectors at terminal states.

\begin{theorem}
\PureNE is in \PTime for terminal-reward games.
\end{theorem}

\section{Randomised Strategies}
\label{sect:randomised}

In this section, we show that the problem \NE is undecidable and,
in fact, not recursively enumerable for turn-based terminal-reward games.
\iffull
The proof proceeds by a reduction from an undecidable problem about
\emph{two-counter machines}.
Such a machine is of the form $\calM=(Q,q_0,\Delta)$, where
\begin{itemize}
\item $Q$ is a finite set of \emph{states},
\item $q_0\in Q$ is the \emph{initial state},
\item $\Delta\subseteq Q\times\upGamma\times Q$ is a set of
\emph{transitions}.
\end{itemize}
The set~$\Gamma$ specifies which instructions $\calM$ may perform on its
counters. For our purposes, the instruction set
$\upGamma\coloneq\{\inc(j),\dec(j),\zero(j):j=1,2\}$ suffices: a counter can
be incremented, decremented, or tested for zero. For $q\in Q$ we
write~$q\Delta$ for the set of all $(\gamma,q')\in\upGamma\times Q$ such that
$(q,\gamma,q')\in\Delta$.
The machine~$\calM$ is \emph{deterministic} if for each $q\in Q$ either
(1)~$q\Delta=\emptyset$, (2)~$q\Delta=\{(\inc(j),q')\}$ for some
$j\in\{1,2\}$ and $q'\in Q$, or
(3)~$q\Delta=\{(\zero(j),q_1),(\dec(j),q_2)\}$ for some $j\in\{1,2\}$ and
$q_1,q_2\in Q$.

A configuration of~$\calM$ is a triple $C=(q,i_1,i_2)\in
Q\times\bbN\times\bbN$, where~$q$ denotes the current state and $i_j$~denotes
the current value of counter~$j$.
A~configuration~$C'=(q',i_1',i_2')$ is a \emph{successor} of
configuration~$C=(q,i_1,i_2)$, denoted by
$C\vdash C'$, if there exists a ``matching'' transition
$(q,\gamma,q')\in\Delta$. For example, $(q,i_1,i_2)\vdash(q',i_1+1,i_2)$ \iff
$(q,\inc(1),q')\in\Delta$. The instruction $\zero(j)$ performs a
\emph{zero test}: $(q,i_1,i_2)\vdash (q',i_1,i_2)$ \iff
$i_1=0$ and $(q,\zero(1),q')\in\Delta$, or $i_2=0$ and
$(q,\zero(2),q')\in\Delta$.

A~\emph{partial computation of~$\calM$} is a sequence
$\rho=\rho(0)\rho(1)\dots$ of configurations such that
$\rho(0)\vdash\rho(1)\vdash\cdots$ and $\rho(0)=(q_0,0,0)$ (the
\emph{initial configuration}). A~partial computation of~$\calM$ is
a \emph{computation of~$\calM$} if it is infinite or it ends in a
configuration~$C$ for which there is no~$C'$ with $C\vdash C'$.
Note that each deterministic two-counter machine has a unique computation.

The \emph{halting problem} is to decide, given a machine~$\calM$,
whether the computation of~$\calM$ is finite. It is well-known that
deterministic two-counter machines are Turing powerful, which makes the
halting problem and its dual, the \emph{non-halting problem}, undecidable,
even when restricted to deterministic two-counter machines. In fact,
the non-halting problem for deterministic two-counter machines is not
recursively enumerable.

To prove the undecidability of \NE, we employ a reduction from the
non-halting problem for deterministic two-counter machines.
More precisely, we show how to compute from such a machine~$\calM$ a game
$(\calG,s_0)$ such that the computation of~$\calM$ is infinite \iff
there exists a Nash equilibrium of $(\calG,s_0)$
where \pl0 receives expected payoff~$\geq 0$. Without loss of generality,
we assume
that in~$\calM$ there is no zero test that is followed
by another zero test: if $(q,\zero(j),q')\in\Delta$, then
$\abs{q'\Delta}\leq 1$.

The game~$\calG$ is played by players $0$, $1$ and 12 other players
$A_j^t$, $B_j^t$, $D^t$ and~$E_j$, indexed by $j\in\{1,2\}$ and $t\in\{0,1\}$.
Intuitively, \pl0 and \pl1 build up the computation of~$\calM$: \pl0 updates
the counters, and \pl1 chooses transitions. Players $A_j^t$ and~$B_j^t$
make sure that \pl0 updates the counters correctly:
players $A_j^0$ and~$A_j^1$ ensure that, in each step, the value of
counter~$j$ is not too high, and players
$B_j^0$ and~$B_j^1$ ensure that, in each step, the value of counter~$j$ is
not too low. More precisely, $A_j^0$ and $B_j^0$ monitor the even steps of the computation, while $A_j^1$ and $B_j^1$ monitor the odd steps.
Finally, players $D^t$ and~$E_j$
ensure that \pl0 uses a randomised strategy of a restricted form.

Let $\upGamma'\coloneq\upGamma\cup\{\init\}$. For each
$q\in Q$, each $\gamma\in\upGamma'$, each $j\in\{1,2\}$ and each
$t\in\{0,1\}$, the game~$\calG$ contains the gadgets $S_{\gamma,q}^t$,
$I_q^t$ and~$C_{\gamma,j}^t$, which are depicted in
\cref{fig:simulation}.
\begin{figure*}
\begin{tikzpicture}[x=1.3cm,y=1.1cm,->]
\tikzset{every label/.style={font=\footnotesize}}
\tikzset{every node/.style={font=\footnotesize}}

\begin{scope} 
\node (caption) at (-0.8,0.9) [anchor=west] {$S_{\gamma,q}^t$:};

\node (1) at (0,0) [play,label=left:$A_1^t$,inner sep=0cm] {$s_{\gamma,q}^t$};
\node (1a) at (1,0) [end,label=right:$\begin{map}0 & -1 \\ A_1^t & \phantom{+}1\end{map}$] {};
\node (2) at (0,-1) [play,label=left:$A_2^t$] {};
\node (2a) at (1,-1) [end,label=right:$\begin{map}0 & -1 \\ A_2^t & \phantom{+}1\end{map}$] {};
\node (3) at (0,-2) [play,label=left:$B_1^t$] {};
\node (3a) at (1,-2) [end,label=right:$\begin{map}0 & -1 \\ B_1^t & \phantom{+}1\end{map}$] {};
\node (4) at (0,-3) [play,label=left:$B_2^t$] {};
\node (4a) at (1,-3) [end,label=right:$\begin{map}0 & -1 \\ B_2^t & \phantom{+}1\end{map}$] {};
\node (5) at (0,-4) [play,label=left:$D^t$] {};
\node (5a) at (1,-4) [end,label={right:$\begin{map}0 & -1 \\ D^t & \phantom{+}2\end{map}$}] {};
\node (55) at (0,-5) [play,label=left:$D^{1-t}$] {};
\node (55a) at (1,-5) [end,label={right:$\begin{map}0 & -1 \\ D^{1-t} & \phantom{+}1\end{map}$}] {};
\node (6) at (0,-6.1) [play,label=left:$0$,fill=black!20] {};
\node (6a) at (1,-6.1) [play,label=above:$E_1$] {};
\node (6b) at (2,-6.1) [play,label=above:$E_2$] {};
\node (6c) at (3,-6.1) [play,label=right:$0$] {};
\node (7) at (0,-7.1) {$I_q^t$};
\node (7a) at (1,-7.1) [end,label=below:$\begin{map}0 & -1 \\ E_1 & \phantom{+}1\end{map}$] {};
\node (7b) at (2,-7.1) [end,label=below:$\begin{map}0 & -1 \\ E_2 & \phantom{+}1\end{map}$] {};
\node (5c) at (3,-5.1) {$C_{\gamma,1}^t$};
\node (7c) at (3,-7.1) {$C_{\gamma,2}^t$};

\draw (0,0.7) to (1);
\draw (1) to (2); \draw (1) to (1a);
\draw (2) to (3); \draw (2) to (2a);
\draw (3) to (4); \draw (3) to (3a);
\draw (4) to (5); \draw (4) to (4a);
\draw (5) to (55); \draw (5) to (5a);
\draw (55) to (6); \draw (55) to (55a);
\draw (6) to (7); \draw (6) to (6a);
\draw (6a) to (6b); \draw (6a) to (7a);
\draw (6b) to (6c); \draw (6b) to (7b);
\draw (6c) to (5c); \draw (6c) to (7c);
\end{scope}

\begin{scope}[yshift=-10.44cm] 
\node  (caption) at (-0.8,0.9) [anchor=west]
 {$I_q^t$ for $q\Delta=\{(\inc(j),q')\}$:};

\node (1) at (0,0) [play,label={above:$1$}] {};
\node (2) at (0.7,0) [coordinate,label={right:$S_{\inc(j),q'}^{1-t}$}] {};

\draw (-0.7,0) to (1);
\draw (1) to (2);
\end{scope}

\begin{scope}[yshift=-12.44cm] 
\node  (caption) at (-0.8,0.9) [anchor=west]
 {$I_q^t$ for $q\Delta=\{(\zero(j),q_1),(\dec(j),q_2)\}$:};

\node (1) at (0,-0.2) [play,label={above:$1$}] {};
\node (2) at (0.8,0.3) [coordinate,label={right:$S_{\zero(j),q_1}^{1-t}$}] {};
\node (3) at (0.8,-0.7) [coordinate,label={right:$S_{\dec(j),q_2}^{1-t}$}] {};

\draw (-0.7,-0.2) to (1);
\draw (1) to (2);
\draw (1) to (3);
\end{scope}

\begin{scope}[yshift=-14.79cm] 
\node (caption) at (-0.8,0.9) [anchor=west]
 {$I_q^t$ for $q\Delta=\emptyset$:};

\node (1) at (0,0.3) [end,label={right:$0\colon -1$}] {};

\draw (-0.7,0.3) to (1);
\end{scope}
\begin{scope}[xshift=6cm,yshift=0cm] 
\node (caption) at (-0.9,0.9) [anchor=west]
 {$C_{\gamma,j}^t$ for $\gamma\notin\{\init,\inc(j),\dec(j),\zero(j)\}$:};

\node (0) at (0,-1.8) [play,label=below:$0$] {};
\node (1) at (1.2,-1.8) [end,fill=black!20,label={right:$\left\{\begin{map} A_j^t & 2 \\ A_j^{1-t} & 2 \\ B_j^t & 2 \\ B_j^{1-t} & 2 \\ D^t & 3 \\ E_j & 2 \end{map}\right.$}] {};
\node (2) at (0,-0.3) [end,label={right:$\left\{\begin{map} A_j^t & 3 \\ B_j^t & 1 \\ B_j^{1-t} & 4 \\ D^t & 3 \\ E_j & 2 \end{map}\right.$}] {};

\draw (-0.7,-1.8) to (0);
\draw (0) to (1);
\draw (0) to (2);
\end{scope}

\begin{scope}[xshift=6cm,yshift=-4.6cm] 
\node (caption) at (-0.9,0.9) [anchor=west]
 {$C_{\gamma,j}^t$ for $\gamma=\inc(j)$:};

\node (0) at (0,-1.8) [play,label=below:$0$] {};
\node (1) at (1.2,-1.8) [end,fill=black!20,label={right:$\left\{\begin{map} A_j^t & 2 \\ A_j^{1-t} & 4 \\ B_j^t & 2 \\ D^t & 3 \\ E_j & 2 \end{map}\right.$}] {};
\node (2) at (0,-0.3) [end,label={right:$\left\{\begin{map} A_j^t & 3 \\ B_j^t & 1 \\ B_j^{1-t} & 4 \\ D^t & 3 \\ E_j & 2 \end{map}\right.$}] {};

\draw (-0.7,-1.8) to (0);
\draw (0) to (1);
\draw (0) to (2);
\end{scope}

\begin{scope}[xshift=6cm,yshift=-9cm] 
\node (caption) at (-0.9,0.9) [anchor=west]
 {$C_{\gamma,j}^t$ for $\gamma=\dec(j)$:};

\node (0) at (0,-1.8) [play,label=below:$0$] {};
\node (1) at (1.2,-1.8) [end,fill=black!20,label={right:$\left\{\begin{map} A_j^t & 2 \\ A_j^{1-t} & 1 \\ B_j^t & 2 \\ B_j^{1-t} & 3 \\ D^t & 3 \\ E_j & 2 \end{map}\right.$}] {};
\node (2) at (0,-0.3) [end,label={right:$\left\{\begin{map} A_j^t & 3 \\ B_j^t & 1 \\ B_j^{1-t} & 4 \\ D^t & 3 \\ E_j & 2 \end{map}\right.$}] {};

\draw (-0.7,-1.8) to (0);
\draw (0) to (1);
\draw (0) to (2);
\end{scope}

\begin{scope}[xshift=6cm,yshift=-13.7cm] 
\node (caption) at (-0.9,0.9) [anchor=west]
 {$C_{\gamma,j}^t$ for $\gamma\in\{\init,\zero(j)\}$:};

\node (0) at (0,-0.7) [play,label=below:$0$] {};
\node (1) at (1.2,-0.7) [end,fill=black!20,label={right:$\left\{\begin{map} 1 & 1 \\ A_j^t & 2 \\ A_j^{1-t} & 2 \\ B_j^t & 2 \\ B_j^{1-t} & 2 \\ D^t & 3 \\ E_j & 2 \end{map}\right.$}] {};

\draw (-0.7,-0.7) to (0);
\draw (0) to (1);
\end{scope}

\end{tikzpicture}
\caption{\label{fig:simulation}Simulating a two-counter machine}
\end{figure*}
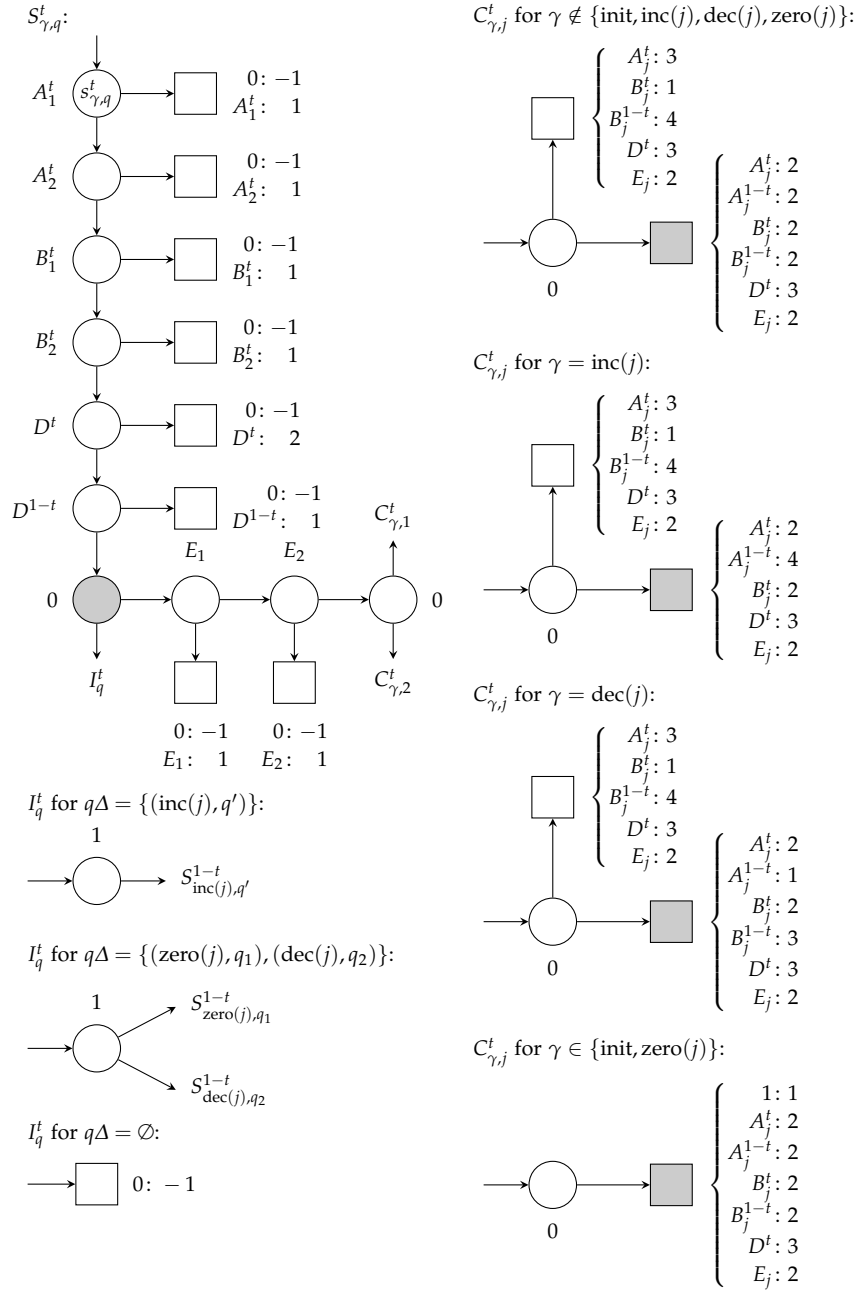
The initial state of~$\calG$ is~$s_0\coloneq s_{\init,q_0}^0$.
Note that in the gadget~$S_{\gamma,q}^t$, each of the players $A_j^t$,
$B_j^t$, $D^t$ and~$E_j$ may unilaterally decide to \emph{quit the game},
which gives the respective player a payoff of $1$ or~$2$, but payoff~$-1$
to player $0$.

It will turn out that \pl1 will play a pure strategy in any Nash equilibrium
of $(\calG,s_0)$ where \pl0 receives expected payoff~$0$, except possibly for
histories that are not consistent with the equilibrium. Moreover, \pl0 has
to play a uniform distribution inside~$S_{\gamma,q}^t$.
Formally, we say that a strategy
profile~$\vec{\sigma}$ of~$\calG$ is \emph{safe} if 1.\
$\sigma_0(xs)$ assigns probability~$\frac{1}{2}$ to both outgoing
transitions for all histories~$xs$
consistent with~$\vec{\sigma}$ and ending in a
state $s\in S_{\gamma,q}^t$ controlled by \pl0,
and 2.\  $\sigma_1(xs)$ is degenerate for all histories~$xs$ consistent
with~$\vec{\sigma}$ and ending in a state~$s$ controlled by \pl1.

For each safe strategy profile~$\vec{\sigma}$ where \pl0
receives expected payoff~$0$, let $x_0s_0\prec x_1s_1\prec x_2s_2\prec \dots$
($x_i\in S^*$, $s_i\in S$, $x_0=\epsilon$) be the unique sequence consisting of
all histories~$xs$ of~$(\calG,s_0)$ consistent with~$\vec{\sigma}$ that
end in a state~$s$ of the form $s=s_{\gamma,q}^t$. This sequence is infinite
because $\vec{\sigma}$ is safe and \pl0 receives expected payoff~$0$.
Additionally, let $q_0,q_1,\dots$ be the corresponding sequence of states and
$\gamma_0,\gamma_1,\dots$ be the corresponding sequence of instructions,
\ie $s_n=s_{\gamma_n,q_n}^0$ or $s_n=s_{\gamma_n,q_n}^1$ for all $n\in\bbN$.
For each $j\in\{1,2\}$ and $n\in\bbN$, we~define two conditional expectations
as follows:
\begin{align*}
a_j^n &\coloneq\Exp_{s_0}^{\vec{\sigma}}(\pay_{A_j^{n\bmod 2}}\mid
 x_n s_n\cdot S^\omega)\,; \\
b_j^n &\coloneq\Exp_{s_0}^{\vec{\sigma}}(\pay_{B_j^{n\bmod 2}}\mid
 x_n s_n\cdot S^\omega)\,.
\end{align*}
Note that at every terminal state of the counter gadgets $C_{\gamma,j}^t$
and~$C_{\gamma,j}^{1-t}$ the rewards of \pl{A_j^t} and \pl{B_j^t} sum up to~$4$.
For each~$j$,
the conditional probability that, given the history~$x_n s_n$, we reach
such a state is $\sum_{k\in\bbN}\frac{1}{2^k}\cdot\frac{1}{4}=\frac{1}{2}$.
Hence, $a_j^n+b_j^n=2$ for all $n\in\bbN$.
We say that $\vec{\sigma}$~is \emph{stable} if $a_j^n=1$ or,
equivalently, $b_j^n=1$ for each $j\in\{1,2\}$ and for all
$n\in\bbN$.

Finally, for each $j\in\{1,2\}$ and $n\in\bbN$, we define a
number~$c_j^n\in [0,1]$ as follows: After the history $x_n s_n$, with
probability~$\frac{1}{4}$ the play proceeds to the state controlled by
\pl0 in the counter gadget~$C_{\gamma_n,j}^{n\bmod 2}$. The number~$c_j^n$ is
defined as the probability that \pl0 plays to the neighbouring
grey state. Note that, by the construction of~$\calG$, it holds that $c_j^n=1$
if $\gamma_n=\zero(j)$ or $\gamma_n=\init$. In particular, $c_1^0=c_2^0=1$.

\begin{lemma}\label{lemma:reduction}
Let $\vec{\sigma}$ be a safe strategy profile with expected payoff~$0$
for \pl0. Then $\vec{\sigma}$~is stable \iff
\begin{equation}\label{eq:counter-update}
c_j^{n+1}=\begin{cases}
  \frac{1}{2}\cdot c_j^n & \text{if $\gamma_{n+1}=\inc(j)$,} \\
  2\cdot c_j^n & \text{if $\gamma_{n+1}=\dec(j)$,} \\
  c_j^n=1 & \text{if $\gamma_{n+1}=\zero(j)$,} \\
  c_j^n & \text{otherwise.}
  \end{cases}
\end{equation}
for each $j\in\{1,2\}$ and for all $n\in\bbN$.
\end{lemma}

To prove the lemma, consider a safe strategy
profile~$\vec{\sigma}$ of~$\calG$ with expected payoff~$0$ for \pl0. For each
$j\in\{1,2\}$ and $n\in\bbN$, we define yet another conditional expectation
\begin{equation*}
p_j^n\coloneq\Exp_{s_0}^{\vec{\sigma}}(\pay_{A_j^{n\bmod 2}}\mid
 x_n s_n\cdot S^\omega\setminus x_{n+2} s_{n+2}\cdot S^\omega)\,.
\end{equation*}
The following claim relates the numbers $a_j^n$ and~$p_j^n$.

\begin{claim*}
Let $j\in\{1,2\}$. Then $a_j^n=1$ for all $n\in\bbN$ \iff
$p_j^n=\frac{3}{4}$ for all $n\in\bbN$.
\end{claim*}
\begin{proof}
($\Rightarrow)$ Assume that $a_j^n=1$ for all $n\in\bbN$.
We have $a_j^n=p_j^n+\frac{1}{4}\cdot a_j^{n+2}$ and therefore
$1=p_j^n+\frac{1}{4}$ for all $n\in\bbN$. Hence,
$p_j^n=\frac{3}{4}$ for all $n\in\bbN$.

($\Leftarrow$) Assume that $p_j^n=\frac{3}{4}$ for all $n\in\bbN$.
Since $a_j^n=p_j^n+\frac{1}{4}\cdot a_j^{n+2}$ for all $n\in\bbN$,
the numbers~$a_j^n$ have to satisfy the following recurrence: $a_j^{n+2}=
4a_j^n-3$. Since all the numbers~$a_j^n$ are bounded by the
minimum and maximum reward for \pl{A_j^{n\bmod 2}},
we have $0\leq a_j^n\leq 4$ for all $n\in\bbN$. It is easy
to see that the only values for $a_j^0$ and~$a_j^1$
such that $0\leq a_j^n\leq 4$ for all $n\in\bbN$ are $a_j^0=a_j^1=1$.
But this implies that $a_j^n=1$ for all $n\in\bbN$.\qed
\end{proof}

\begin{proof}[of \cref{lemma:reduction}]
By the previous claim, it suffices to show that $p_j^n=\frac{3}{4}$
\iff \cref{eq:counter-update} holds.
Let $j\in\{1,2\}$, $n\in\bbN$ and $t=n\bmod 2$. The number~$p_j^n$ can
be expressed as a weighted average of the expected payoff for \pl{A_j^t}
inside~$C_{\gamma_n,j}^t$ and the expected payoff for \pl{A_j^t}
inside~$C_{\gamma_{n+1},j}^{1-t}$.
The first payoff does not depend on~$\gamma_n$, but the second depends
on~$\gamma_{n+1}$. Let us consider the case that $\gamma_{n+1}=\inc(j)$. In
this case, $p_j^n$~equals
\[\tfrac{1}{4}\cdot\big(c_j^n\cdot 2+(1- c_j^n)\cdot 3\big)
+\tfrac{1}{8}\cdot c_j^{n+1}\cdot 4
=\tfrac{3}{4}-\tfrac{1}{4}\cdot c_j^n+\tfrac{1}{2}\cdot c_j^{n+1}.\]
Obviously, this sum equals~$\frac{3}{4}$ \iff
$c_j^{n+1}=\frac{1}{2}\cdot c_j^n$. For any other value
of~$\gamma_{n+1}$, the argumentation is similar.\qed
\end{proof}

The next lemma states that every Nash equilibrium with expected payoff~$0$
for \pl0 is, in fact, safe.

\begin{lemma}\label{lemma:ne-safe}
Let $\vec{\sigma}$ be a Nash equilibrium of $(\calG,s_0)$ with expected
payoff~$0$ for \pl0. Then $\vec{\sigma}$~is safe.
\end{lemma}

\begin{proof}
We start by proving that \pl0 plays a uniform distribution
inside~$S_{\gamma,q}^t$. We prove this separately for histories that end in a
white state and histories that end in a grey state.

Let $xs$ be a history consistent with~$\vec{\sigma}$ and ending
in a white state $s\in S_{\gamma,q}^t$ controlled by \pl0.
Since the players~$E_1$
and~$E_2$ can ensure payoff~$1$ by quitting the game,
\pl0 has to play to $C_{\gamma,1}^t$
and~$C_{\gamma,2}^t$ with probability~$\frac{1}{2}$ each. Otherwise,
$\vec{\sigma}$~would not be a Nash equilibrium.
	
Now let $xs$ be a history consistent with~$\vec{\sigma}$ and ending
in a grey state $s\in S_{\gamma,q}^t$ controlled by \pl0.
In the following, let $t=0$; the proof for $t=1$ is analogous.
Denote by~$p$ the probability that \pl0 plays to~$t\in I_q^t$
after the history~$xs$. For $i\in\{0,1\}$, let
\[d^i=\Exp_{s_0}^{\vec{\sigma}}(\pay_{D^i}\mid xst\cdot S^\omega).\]
By the definition of the game, we have $d^0\geq 1$ and $d^1\geq 2$.
On the other hand, since at every terminal state the sum of the rewards for
players $D^0$ and~$D^1$ is at most~$3$, we have $d^0+d^1\leq 3$. Hence,
$d^0=1$ and $d^1=2$. Consider the expected payoffs for players $D^0$
and~$D^1$ after the history~$xs$:
\begin{align*}
\Exp_{s_0}^{\vec{\sigma}}(\pay_{D^0}\mid xs\cdot S^\omega)
&=(1-p)\cdot 3+p\cdot d^0=3-2p; \\
\Exp_{s_0}^{\vec{\sigma}}(\pay_{D^1}\mid xs\cdot S^\omega)
&=p\cdot d^1=2p\,.
\end{align*}
Since $\vec{\sigma}$~is a Nash equilibrium, these numbers are bounded from
below by $2$ and~$1$, respectively (otherwise, it would be better for player
$D^0$ or~$D^1$ to quit the game). Hence, $p=\frac{1}{2}$.

To prove that $\vec{\sigma}$~is safe, it remains to be shown that
\pl1 plays a degenerate distribution for all
histories~$xs$ consistent with~$\vec{\sigma}$ and ending in a state
$s\in I_q^t$.
Towards a contradiction, assume that $xs$~is such a history and
that $\sigma_1(xs)$~assigns probability~$>0$ to two distinct successor
states. Hence, $q\Delta=\{(\zero(j),q_1),(\dec(j),q_2)\}$ for some
$j\in\{1,2\}$ and $q_1,q_2\in Q$. By our assumption that there are no
consecutive zero tests and since \pl0 receives expected payoff~$0$,
\begin{align*}
\Exp_{s_0}^{\vec{\sigma}}(\pay_1\mid xs\cdot s_{\zero(j),q_1}^{1-t}
\cdot S^\omega) &\geq \tfrac{1}{4}\,, \\
\shortintertext{but}
\Exp_{s_0}^{\vec{\sigma}}(\pay_1\mid xs\cdot s_{\dec(j),q_2}^{1-t}
\cdot S^\omega) &\leq \tfrac{1}{6}\,.
\end{align*}
Hence, \pl1 could improve her payoff by playing to~$s_{\zero(j),q_1}^{1-t}$
with probability~$1$, a contradiction to $\vec{\sigma}$~being a Nash
equilibrium.\qed
\end{proof}

\noindent
Finally, we can prove the following theorem.
\fi

\begin{theorem}\label{thm:undecidability}
\NE is not recursively enumerable, even for turn-based 14-player
terminal-reward games.
\end{theorem}

\iffull
\begin{proof}
We claim that the function mapping a deterministic two-counter machine~$\calM$
to the 14-player game $(\calG,s_0)$ as described above realises a many-one
reduction from the non-halting problem to \NE. Clearly, $\calG$~can be computed
from~$\calM$. We prove that the computation of~$\calM$ is infinite \iff
$(\calG,s_0)$ has a Nash equilibrium in which \pl0 receives expected
payoff (at least)~$0$.

($\Rightarrow$) Assume that the computation $\rho=\rho(0)\rho(1)\dots$
of~$\calM$ is infinite. \Pl0's equilibrium strategy~$\sigma_0$ can be described
as follows: For a history that ends at the unique state controlled by \pl0
in the gadget~$C_{\gamma,j}^t$ after visiting a state of the form
$s_{\gamma',q}^t$ or~$s_{\gamma',q}^{1-t}$ exactly $n>0$ times, \pl0 plays
to the grey successor state with probability~$2^{-i}$, where $i$~is the value
of counter~$j$ in configuration ${\rho(n-1)}$. Moreover, for a history that
ends at a state controlled by \pl0 in the gadget~$S^t_{\gamma,q}$, \pl0 plays
to both successors with probability~$\frac{1}{2}$ each.

The only place where \pl1 has a choice is the sole state in the gadget~$I_q^t$
for $q\Delta=\{(\zero(j),q_1),(\dec(j),q_2)\}$. If the play arrives at such a
state after visiting a state of the form $s_{\gamma,q'}^t$
or~$s_{\gamma,q'}^{1-t}$ exactly $n>0$ times, then \pl1's pure
strategy~$\sigma_1$ prescribes to play to~$S_{\zero(j),q_1}^{1-t}$ if the value
of counter~$j$ in configuration~$\rho(n-1)$ is zero and
to~$S_{\dec(j),q_2}^{1-t}$ if the value of counter~$j$ in
configuration~$\rho(n-1)$ is non-zero.

Any other player's pure strategy is defined as follows: After a history
ending in~$S_{\gamma,q}^t$, the strategy prescribes to quit the game \iff
the history is not compatible with~$\rho$ (\ie the corresponding sequence of
instructions does not match~$\rho$).

Note that the resulting strategy profile~$\vec{\sigma}$ is
safe. Moreover, since \pl0 and \pl1 follow the computation
of~$\calM$, a terminal state inside one of the counter
gadgets~$C_{\gamma,j}^t$ is reached with probability~$1$. Since \pl0 receives
reward~$0$ at any such terminal state, \pl0's expected payoff equals~$0$.
Finally, by the definition of~$\vec{\sigma}$, for each $j\in\{1,2\}$ and
for all $n\in\bbN$, if $i$ and~$i'$ are the values of counter~$j$ in
configuration~$\rho(n)$ and configuration~$\rho(n+1)$, respectively, then
$c_j^n=2^{-i}$, $c_j^{n+1}=2^{-i'}$, and $\gamma_{n+1}$~is the instruction
corresponding to the counter update from $\rho(n)$ to~$\rho(n+1)$. Hence,
\cref{eq:counter-update} holds, and we can conclude from \cref{lemma:reduction}
that $\vec{\sigma}$~is stable.

We claim that $\vec{\sigma}$~is, in fact, a Nash equilibrium of $(\calG,s_0)$:
It is obvious that \pl0 cannot improve her payoff. If \pl1 deviates, then with
positive probability we reach a history that is not compatible with~$\rho$;
hence, player $A_1^0$ or~$A_1^1$ will quit the game, which ensures that \pl1
will receive payoff~$0$ after this history. Since $\vec{\sigma}$~is
stable, none of the players $A_j^t$ or~$B_j^t$ can improve her payoff. Finally,
the expected payoffs of \pl{D^t} and \pl{D^{1-t}} from $s^t_{\gamma,q}$ equal
$2$ and~$1$, respectively, which is the same as they would get if they quit
the game. The reasoning for players $E_1$ and~$E_2$ is analogous.

($\Leftarrow$) Assume that $\vec{\sigma}$~is a Nash equilibrium of
$(\calG,s_0)$ with expected payoff $\geq 0$ for \pl0. Since $0$~is the
maximum reward for \pl0, this means that the expected payoff
of~$\vec{\sigma}$ for \pl0 equals~$0$. From \cref{lemma:ne-safe}, we
can conclude that $\vec{\sigma}$~is safe.
To apply \cref{lemma:reduction} and obtain \cref{eq:counter-update}, it
remains to be shown that $\vec{\sigma}$~is stable. In order to derive a
contradiction, assume that there exists $j\in\{1,2\}$ and $n\in\bbN$ such that
either $a_j^n<1$ or $a_j^n>1$, \ie $b_j^n<1$. In the first case,
\pl{A_j^{n\bmod 2}} could improve
her payoff by quitting the game after history~$x_n s_n$, while in the second
case, \pl{B_j^{n\bmod 2}} could improve her payoff by quitting the game, again
a contradiction to $\vec{\sigma}$~being a Nash equilibrium.

From \cref{eq:counter-update} and the fact that $c_j^0=1$, it follows that
each~$c_j^n$ is of the form $c_j^n=2^{-i}$ with $i\in\bbN$. We denote
by~$i_j^n$ the unique number~$i$ such that $c_j^n=2^{-i}$ and set
$\rho(n)=(q_n,i_1^n,i_2^n)$ for each $n\in\bbN$. We claim that
$\rho\coloneq\rho(0)\rho(1)\dots$ is in fact the computation of~$\calM$.
In~particular, this computation is infinite. It suffices to verify the
following two properties:
\begin{itemize}
 \item $\rho(0)=(q_0,0,0)$.
 \item $\rho(n)\vdash\rho(n+1)$ for all $n\in\bbN$.
\end{itemize}
The first property is immediate. To prove the second property, let
$\rho(n)=(q,i_1,i_2)$ and $\rho(n+1)=(q',i_1',i_2')$. Hence, $s_n$~lies
inside~$S_{\gamma,q}^t$, and $s_{n+1}$~lies inside~$S_{\gamma',q'}^{1-t}$ for
suitable $\gamma,\gamma'$ and $t=n\bmod 2$.
We only prove the claim for $q\Delta=\{(\zero(1),q_1),(\dec(1),q_2)\}$; the
other cases are similar. Note that, by the construction of the
gadget~$I_q^t$, it must be the case that either $q'=q_1$ and
$\gamma'=\zero(1)$, or $q'=q_2$ and $\gamma'=\dec(1)$. By
\cref{eq:counter-update}, if $\gamma'=\zero(1)$, then $i_1'=i_1=0$ and
$i_2'=i_2$, and if $\gamma'=\dec(1)$, then $i_1'=i_1-1$ and $i_2'=i_2$. This
implies $\rho(n)\vdash\rho(n+1)$: On the one hand, if $i_1=0$, then
$i_1'\neq i_1-1$, which implies $\gamma'\neq\dec(1)$ and thus
$\gamma'=\zero(1)$, $q'=q_1$ and $i_1'=i_1=0$. On the other hand, if $i_1>0$,
then $\gamma'\neq\zero(1)$ and thus $\gamma'=\dec(1)$, $q'=q_2$ and
$i_1'=i_1-1$.\qed
\end{proof}
\fi

\iffull
For games that are not turn-based, we can show the stronger theorem that
the set of all games that have a Nash equilibrium is
not recursively enumerable.
\fi

\begin{corollary}
\label{thm:undecidability-2}
The set of all initialised concurrent 14-player terminal-reward games that
have a Nash equilibrium is not recursively enumerable.
\end{corollary}

\iffull
\begin{proof}
The proof is analogous to the proof of \cref{cor:statnene-np-complete},
but we use the game~$\calG_1$ from \cref{ex:hide-or-run} instead of
the game~$\calG_2$, and
we set the reward for \pl0 in each state of~$\calG_1$ to~$0$.\qed
\end{proof}
\fi

\section{Conclusion}

We have analysed the complexity of Nash equilibria in concurrent games with
limit-average objectives. In particular, we have shown that randomisation in
strategies leads to undecidability, while restricting to pure strategies
retains decidability. This is in contrast to stochastic games, where pure
strategies lead to undecidability~\cite{UmmelsW09}. While we have provided
matching and lower bounds in most cases, there remain some problems
where we do not know the exact complexity. Apart from \StatNE, these
include the problem \PureNE when restricted to a bounded number of players.

\bibliographystyle{arxiv}
\bibliography{all}

\section*{Appendix}

This appendix is devoted to the proof of \cref{thm:mp-path}, which is
restated here.

\begin{reptheorem}{thm:mp-path}
Given a finite directed graph $G=(V,E)$ with weight functions
$r_0,\dots,r_{k-1}\colon V\to\bbQ$, $v_0\in V$, and
$\vec{x},\vec{y}\in(\bbQ\cup\{\pm\infty\})^k$, we can decide in polynomial
time whether there exists an infinite path~$\pi=v_0 v_1\ldots$ in~$G$ with
$x_i\leq\liminf_{n\to\infty}\frac{1}{n}\sum_{j=0}^{n-1}r_i(v_j)\leq y_i$
for all $i=0,\dots,k-1$.
\end{reptheorem}

In the following, let $G=(V,E)$ be a finite directed graph with weight
functions $r_0,\dots,r_{k-1}\colon V\to\bbQ$, and set $\Ind=\{0,1,\dots,k-1\}$.
Given a vertex $v\in V$, we write
$\In(v)$ and $\Out(v)$ for the set of all edges that end, respectively
start, in~$v$. Moreover, given an edge $e=(u,v)\in E$
we set $r_i(e)\coloneq r_i(u)$.
We extend the weight functions~$r_i$ to finite paths by setting
$r_i(v_1\ldots v_n)=\sum_{j=1}^n r_i(v_j)$. If $\pi=\pi(0)\pi(1)\ldots$
is an infinite path and $n\in\bbN$, we write $\pi\restrict n$ for
the finite path $\pi(0)\ldots\pi(n-1)$, and we set
$\pay_i(\pi)\coloneq\liminf_{n\to\infty}r_i(\pi\restrict n)/n$, \ie
$\pay_i(\pi)$~is precisely the limit-average weight of the path~$\pi$
\wrt the weight function~$r_i$. Finally, $\pay(\pi)$~denotes the vector
$(\pay_i(\pi))_{i\in\Ind}$.
Now consider the following linear constraints
over the variables $f_{i,e}$, where $i\in\Ind$ and $e\in E$:
\begin{enumerate}[label=(\arabic*)]
\item\label{const:positive} $f_{i,e}\geq 0$ for all $i\in\Ind$
and $e\in E$;
\item\label{const:sum} $\sum_{e\in E}f_{i,e}=1$ for all $i\in\Ind$;
\item\label{const:balance} $\sum_{e\in\In(v)} f_{i,e}=
 \sum_{e\in\Out(v)} f_{i,e}$ for all $i\in\Ind$ and $v\in V$;
\item\label{const:thresholds} $x_i\leq\sum_{e\in E}
 f_{i,e}\cdot r_i(e)\leq y_i$ for all $i\in\Ind$;
\item\label{const:bound} $\sum_{e\in E} f_{i,e}\cdot r_i(e)\leq
 \sum_{e\in E} f_{j,e}\cdot r_i(e)$ for all $i,j\in\Ind$.
\end{enumerate}

\begin{lemma}\label{lemma:sound}
If there exists an infinite path~$\pi$ in~$G$ such that
$\vec{x}\leq\pay(\pi)\leq\vec{y}$, then there exists a solution to
\ref{const:positive}--\ref{const:bound}.
\end{lemma}

\begin{proof}
Let $\pi=\pi(0)\pi(1)\ldots$ be an infinite path in~$G$ such that
$\vec{x}\leq\pay(\pi)\leq\vec{y}$.
Given $n\in\bbN$ and $e\in E$, define
$\kappa(n,e)\coloneq\abs{\{j<n:(\pi(j),\pi(j+1))=e\}}$. Moreover,
for $n>0$, set $\lambda(n,e)=\kappa(n,e)/n$.
Note that $0\leq\lambda(n,e)\leq 1$ for all $e\in E$ and $n\in\bbN$.
In order to define the numbers $f_{i,e}$, let us now fix $i\in\Ind$.
Since $\pay_i(\pi)=\liminf_{n\to\infty} r_i(\pi\restrict n)/n$,
there exist natural numbers $0<k_0^i<k_1^i<\cdots$ such that
$\pay_i(\pi)=\lim_{n\to\infty} r_i(\pi\restrict k_n^i)/k_n^i$.
Now we define a sequence
$\phi_0^i,\phi_1^i,\ldots$ of vectors $\phi_n^i\in\bbR^E$ by setting $\phi_n^i(e)=\lambda(k_n^i,e)$. Since this sequence is bounded,
by the Bolzano-Weierstrass theorem, there exists a converging
subsequence $\psi_0^i,\psi_1^i,\ldots$ of this sequence.
We set $f_{i,e}=\lim_{n\to\infty}\psi^i_n(e)$ for all $e\in E$.

We claim that the numbers $(f_{i,e})_{i\in\Ind,e\in E}$ form a solution of
\ref{const:positive}--\ref{const:bound}. That \ref{const:positive} holds is
obvious from the definition. \ref{const:sum} follows from the fact that
$\sum_{e\in E}\lambda(n,e)=1$ for all $n\in\bbN$. To show that
\ref{const:balance} holds, fix $v\in V$. Note that we have
$\sum_{e\in\In(v)}\kappa(n,e)-\sum_{e\in\Out(v)}\kappa(n,e)\in\{-1,0,1\}$
and therefore
$-1/n\leq\sum_{e\in\In(v)}\lambda(n,e)-\sum_{e\in\Out(v)}\lambda(n,e)\leq 1/n$
for all $n\in\bbN$. Hence,
the terms $\sum_{e\in\In(v)}\phi^i_n(e)-\sum_{e\in\Out(v)}\phi^i_n(e)$
converge to~$0$ when $n$~goes to infinity.
Since $\psi_0^i,\psi_1^i,\ldots$ is a subsequence of
$\phi_0^i,\phi_1^i,\ldots$, the same is true for the terms
$\sum_{e\in\In(v)}\psi_n^i(e)-\sum_{e\in\Out(v)}\psi_n^i(e)$.
Since $\lim_{n\to\infty}\psi_n^i(e)=f_{i,e}$ exists for all $e\in E$,
this implies that $\sum_{e\in\In(v)}f_{i,e}-\sum_{e\in\Out(v)}f_{i,e}=0$,
which proves \ref{const:balance}.
In order to prove \ref{const:thresholds} and \ref{const:bound}, note that
for all $i,j\in\Ind$ we have
\begin{align*}
\pay_i(\pi) &= \liminf_{n\to\infty} r_i(\pi\restrict n)/n \\
&\leq\liminf_{n\to\infty} r_i(\pi\restrict k_n^j)/k_n^j \\
&=\liminf_{n\to\infty}\sum_{e\in E}\lambda(k_n^j,e)\cdot r_i(e) \\
&=\liminf_{n\to\infty}\sum_{e\in E}\phi_n^j(e)\cdot r_i(e) \\
&\leq\lim_{n\to\infty}\sum_{e\in E}\psi_n^j(e)\cdot r_i(e) \\
&=\sum_{e\in E} f_{j,e}\cdot r_i(e)\,.
\end{align*}
Moreover, if $i=j$, both inequalities are
equalities since $\lim_{n\to\infty} r_i(\pi\restrict k_n^i)/k_n^i$ exists
and equals~$\pay_i(\pi)$. Hence, $\sum_{e\in E} f_{i,e}\cdot r_i(e)=\pay_i(\pi)
\leq \sum_{e\in E} f_{j,e}\cdot r_i(e)$ for all $i,j\in\Ind$, which
proves \ref{const:bound}. Finally, \ref{const:thresholds} follows from
the assumption that $\vec{x}\leq\pay(\pi)\leq\vec{y}$.\qed
\end{proof}

\begin{lemma}\label{lemma:factorial}
For all $n\in\bbN$,
\[(n-2)\cdot\sum_{j=1}^{n-1} j!<n!\]
\end{lemma}
\begin{proof}
By induction over~$n$.\qed
\end{proof}

\begin{lemma}\label{lemma:complete}
Assume that $G$~is strongly connected and that there exists a solution to
\ref{const:positive}--\ref{const:bound}.
Then there exists an infinite path~$\pi$ in~$G$ such that
$\vec{x}\leq\pay(\pi)\leq\vec{y}$.
\end{lemma}

\begin{proof}
Let $G$ be strongly connected and assume that there exists a solution
to \ref{const:positive}--\ref{const:bound}. It is well-known that if a
given system of linear constraints has a solution, then there exists one
in rational numbers. Let $(f_{i,e})_{i\in\Ind,e\in E}$ be such a solution,
where \wlg $f_{i,e}=c_{i,e}/d$ with $c_{i,e}\in\bbN$ and
$d\in\bbN\setminus\{0\}$. Finally, let $\vec{z}\in\bbR^\Ind$ be defined by
$z_i=\sum_{e\in E} f_{i,e}\cdot r_i(e)$; by \ref{const:thresholds},
$\vec{x}\leq\vec{z}\leq\vec{y}$. We claim that there exists an
infinite path~$\pi$ in~$G$ with $\pay(\pi)=\vec{z}$.

For each $i\in\Ind$ consider the directed multigraph~$G_i$,
which is derived from~$G$ by replacing a single edge $(u,v)\in E$
by as many as~$c_{i,e}$~edges from $u$ to~$v$. By \ref{const:balance},
we have $\sum_{e\in\In(v)} c_{i,e}=\sum_{e\in\Out(v)} c_{i,e}$ for
all $v\in V$. Hence, in~$G_i$ each vertex has as many incoming edges
as outgoing edges, which is a necessary and sufficient condition for
the existence of an Eulerian cycle in each of the connected
components of~$G_i$. These cycles give rise to (disjoint, not
necessarily simple) cycles $\gamma_1^i,\dots,\gamma_m^i$ in~$G$,
where $m\leq\abs{V}$.

Consider for each $n\in\bbN$ the cycle $\zeta_n^i$ that starts
by repeating the cycle~$\gamma_1^i$ $n$~times, then takes the
shortest path to the first vertex in the cycle~$\gamma_2^i$,
repeats this cycle $n$~times, and so on, until, after repeating
the cycle $\gamma_m^i$ $n$~times, taking the shortest path back
to~$\gamma_1^i$.
Let $M=\max_{v\in V} r_j(v)$ be the maximum weight \wrt~$r_j$. Note that:
\begin{gather*}
n\cdot\sum_{e\in E} c_{i,e}\cdot r_j(e) \leq r_j(\zeta_n^i)
\leq n\cdot\sum_{e\in E}c_{i,e}\cdot r_j(e)+\abs{V}^2\cdot M,\\
n\cdot\sum_{e\in E} c_{i,e} \leq\abs{\zeta_n^i}
\leq n\cdot\sum_{e\in E} c_{i,e}+\abs{V}^2.
\end{gather*}
Hence,
\[
\lim_{n\to\infty} \frac{r_j(\zeta_n^i)}{\abs{\zeta_n^i}}
= \frac{\sum_{e\in E} c_{i,e}\cdot r_j(e)}{\sum_{e\in E} c_{i,e}}
= \frac{\sum_{e\in E} f_{i,e}\cdot r_j(e)}{\sum_{e\in E} f_{i,e}}
= \sum_{e\in E} f_{i,e}\cdot r_j(e)
\geq z_j,\]
where the last inequality follows from \ref{const:bound}. Moreover,
if $i=j$, we have equality, \ie $\lim_{n\to\infty} r_i(\zeta_n^i)/\abs{\zeta_n^i}=z_i$.

The desired infinite path~$\pi$ is the concatenation of finite paths~$\pi_n$,
where $n=1,2\ldots$. The path~$\pi_n$ repeats the cycle
$\zeta_n^{n\bmod k}$ $n!$~times and then takes the shortest path to
the first state on the cycle $\zeta_n^{(n+1)\bmod k}$. We will now
prove that $\pay_0(\pi)=z_0$; for all other weight functions, the proof is
analogous.
For all $n\in\bbN$, we have:
\begin{gather*}
\sum_{j=1}^{nk}j!\,r_0\big(\zeta_j^{j\bmod k}\big)-nk\abs{V} M
\leq r_0(\pi_1\cdots \pi_{nk})
\leq\sum_{j=1}^{nk}j!\,r_0\big(\zeta_j^{j\bmod k}\big)+nk\abs{V} M,\\
\sum_{j=1}^{nk}j!\,\big\lvert\zeta_j^{j\bmod k}\big\rvert
\leq \abs{\pi_1\cdots \pi_{nk}}
\leq \sum_{j=1}^{nk}j!\,\big\lvert\zeta_j^{j\bmod k}\big\rvert+nk\abs{V}\,.
\end{gather*}
By \cref{lemma:factorial}, we have $\lim_{n\to\infty}
\sum_{j=1}^{nk-1}j!/(nk)!=0$.
Hence, and since
$\abs{\zeta_j^i},r_0(\zeta_j^i)\leq j\cdot c$
for some constant~$c$, we have:
\begin{gather*}
\lim_{n\to\infty}
\frac{1}{nk(nk)!}\cdot\sum_{j=1}^{nk}j!\,r_0\big(\zeta_j^{j\bmod k}\big)
=\lim_{n\to\infty}\frac{1}{nk}\cdot r_0(\zeta_{nk}^0)
=\sum_{j=1}^m r_0(\gamma_j^0), \\
\lim_{n\to\infty}
\frac{1}{nk(nk)!}\cdot\sum_{j=1}^{nk}j!\,\big\lvert\zeta_j^{j\bmod k}\big\rvert
=\lim_{n\to\infty}\frac{1}{nk}\cdot\abs{\zeta_{nk}^0}
=\sum_{j=1}^m\abs{\gamma_j^0}\,.
\end{gather*}
Hence,
\[
\lim_{n\to\infty}\frac{r_0(\pi_1\cdots\pi_{nk})}{\abs{\pi_1\cdots\pi_{nk}}}
= \lim_{n\to\infty}\frac{r_0(\zeta^0_{nk})}{\abs{\zeta_{nk}^0}}=z_0\,.\]
We have thus found a subsequence of
$r_0(\pi\restrict n)/n$ that converges to~$z_0$, which implies that
$\pay_0(\pi)=\liminf_{n\to\infty} r_0(\pi\restrict n)/n\leq z_0$.
On the other hand, using the fact that
$\lim_{n\to\infty} r_0(\zeta_n^i)/\abs{\zeta_n^i}\geq z_0$ for
all $i\in\Ind$, we can show that $\pay_0(\pi)\geq z_0$.\qed
\end{proof}

\begin{proof}[of \cref{thm:mp-path}]
Since the limit-average criterion is prefix-independent, it~suffices
to decompose~$G$ into its strongly connected components (which can be
done in linear time) and check for each component~$C$ that is reachable
from~$v_0$ whether exists an infinite path in~$C$ with
$\vec{x}\leq\pay(\pi)\leq\vec{y}$. By \cref{lemma:sound,lemma:complete},
such a path exists \iff there exists a solution to the linear constraints
\ref{const:positive}--\ref{const:bound} derived from~$C$. The
existence of such a solution can be checked in polynomial time
(see~\cite{Schrijver98}).\qed
\end{proof}

\end{document}